\documentclass[3p]{elsarticle}

\usepackage{aas_macros,mathtools,amsmath,amssymb,CJK,lineno,hyperref,epigraph,textcomp,gensymb,color}

\modulolinenumbers[2]
   
\newcommand{\zcorr}[1]{\color{black}#1\color{black}}

\begin{document}

\begin{frontmatter}

\title{Meteor Shower Modeling: Past and Future Draconid Outbursts}


\author[a,b,c]{A. Egal\corref{correspondingauthor}}
\cortext[correspondingauthor]{Corresponding author}
\ead{aegal@uwo.ca}

\author[a,b]{P. Wiegert}
\author[a,b]{P. G. Brown}
\author[d]{D. E. Moser}
\author[a,b]{M. Campbell-Brown}
\author[e]{A. Moorhead}
\author[f]{\zcorr{S. Ehlert}}
\author[e,g]{N. Moticska}


\address[a]{Department of Physics and Astronomy, The University of Western Ontario, London, Ontario N6A 3K7, Canada}
\address[b]{Centre for Planetary Science and Exploration, The University of Western Ontario, London, Ontario N6A 5B8, Canada}
\address[c]{IMCCE, Observatoire de Paris, PSL Research University, CNRS, Sorbonne
Universit\'{e}s, UPMC Univ. Paris 06, Univ. Lille}
\address[d]{Jacobs Space Exploration Group, NASA Meteoroid Environment Office, Marshall Space Flight Center, Huntsville, AL 35812 USA}
\address[e]{NASA Meteoroid Environment Office, Marshall Space Flight Center, Huntsville, AL 35812 USA}
\address[f]{\zcorr{Qualis Corporation, Jacobs Space Exploration Group, NASA Meteoroid Environment Office, Marshall Space Flight Center, Huntsville, AL 35812 USA}}
\address[g]{Embry-Riddle Aeronautical Univeristy, Daytona Beach, FL, USA}

\begin{abstract}
This work presents numerical simulations of meteoroid streams released by comet 21P/Giacobini-Zinner over the period 1850-2030. The initial methodology, based on \cite{Vaubaillon2005}, has been updated and modified to account for the evolution of the comet's dust production along its orbit. The peak time, intensity, and duration of the shower were assessed using simulated activity profiles that are calibrated to match observations of historic Draconid outbursts. The characteristics of all the main apparitions of the shower are reproduced, with a peak time accuracy of half an hour and an intensity estimate correct to within a factor of 2 (visual showers) or 3 (radio outbursts). Our model also revealed the existence of a previously unreported strong radio outburst on October 9 1999, that has since been confirmed by archival radar measurements. The first results of the model, presented in \cite{Egal2018}, provided one of the best predictions of the recent 2018 outburst. Three future radio outbursts are predicted in the next decade, in 2019, 2025 and 2029. The strongest activity is expected in 2025 when the Earth encounters the young 2012 trail. Because of the dynamical uncertainties associated with comet 21P's orbital evolution between the 1959 and 1965 apparitions, observations of the 2019 radio outburst would be particularly helpful to improve the confidence of subsequent forecasts.
\end{abstract}

\begin{keyword}
Meteors \sep Comet Giacobini-Zinner, dust, dynamics
\end{keyword}

\end{frontmatter}



\section{Introduction} \label{intro}

The October Draconid shower (009 DRA) is an autumnal meteor shower known to episodically appear around the 9$^{th}$ of October since 1926. The Draconid shower, with apparitions irregular in time and intensity, has challenged the forecasts of modelers since its discovery. Its parent body is the Jupiter-family comet 21P/Giacobini-Zinner, discovered in 1900, which has an erratic  and highly perturbed orbit \citep{Marsden1971}. The Draconid annual activity is usually barely perceptible (with a rate of a few visual meteors per hour), but the stream occasionally produces strong outbursts and storms (up to ten thousand meteors per hour) that are not directly correlated with the past geometrical configuration between the Earth and the parent comet \citep{Egal2018}. In addition, some showers were observed by naked-eye witnesses or using video devices, while other outbursts (e.g. 2012) were caused by small meteoroids only detectable by radar instruments \citep{Ye2014}.

 Previous attempts to predict upcoming Draconid activity have had mixed success. Of the nine historic outbursts of the shower (1926, 1933, 1946, 1952, 1985, 1998, 2005, 2011 and 2012), five were expected (1926, 1946, 1985, 1998 and 2011) \zcorr{and predictions of three events } correctly estimated the peak time with an accuracy better than two hours. In 1946 the maximum activity occurred close to the descending node of the comet, simplifying timing estimates. \zcorr{\cite{Reznikov1993} accurately predicted the shower return in 1998, caused by an encounter with the meteoroid stream ejected in 1926. More recently, a successful peak time prediction concerned the 2011 outburst. } Numerical simulations of meteoroid streams ejected from 21P conducted by different authors estimated a maximum activity around 20h UT and a secondary peak around 17h on October 8 2011 \citep{Watanabe2008,Vaubaillon2011}; the existence of both peaks was indeed confirmed by independent observation campaigns \cite[e.g.][]{McBeath2012,Trigo2013,Kac2015}.
 
 If different models provided an accurate prediction of the 2011 apparition date, they also led to the first reasonable estimate of the outburst's strength. The activity of a shower, usually characterized by the zenithal hourly rate (ZHR) parameter (i.e. the number of meteors a single observer would see during an hour under ideal conditions), is a quantity that is hard to foresee. To our knowledge, no real intensity predictions were performed for Draconid outbursts prior to the 1998 apparition. The first attempt of \cite{Kresak1993} to estimate the strength of the shower outburst in 1998, expected to be lower than the one occuring in 1985, underestimated the 1998 shower's intensity by a factor of ten. 
 
 In 2011, ZHR estimates of the main peak ranged from 40-50 \citep{Maslov2011} to 600 \citep{Watanabe2008,Vaubaillon2011}, with intermediate activity predictions reaching storm level \citep[ZHR of 7000 in][]{Sigismondi2011}. Numerous observations of the 2011 Draconids estimated a ZHR between 300 \citep{Kero2012,McBeath2012} and 400-460 \citep{Trigo2013,Kac2015}. Among the wide range of ZHR predictions for 2011 published in the literature, some were in good agreement with the observations \citep{Watanabe2008, Vaubaillon2011}. This success could have settled the question of Draconid forecasts using numerical simulations. However, a completely unexpected radio storm reaching a ZHR of 9000$\pm$1000 meteors per hour was observed one year later \citep{Ye2014}, highlighting the need to improve the intensity predictions of meteor showers. 
 
The motivation underlying the present work was therefore to be able to predict not only the date and time of appearance of a Draconid shower, but to also provide a quantitative estimate of its intensity with relative confidence. Since the numerical modeling of meteoroid streams ejected from the parent comet is the only efficient tool to predict such a shower, we performed simulations of the \{21P, Draconid\} complex using the model of \cite{Vaubaillon2005}. For the first time, we simulated ZHR profiles for each year of predicted enhanced Draconid activity and compared them with the available observations. The simulated profiles were then calibrated against observations of the timing, activity profile and strength of historic Draconid outbursts to reinforce the reliability of future forecasts. This work presents the implementation and results of our meteoroid stream modeling, applied specifically to the Draconid shower. The structure of the paper is comprised of four main parts, some of which consist of multiple sections. 
 
\begin{itemize}
 \item  In Section \ref{sec:observations}, we detail the main characteristics of the historic Draconid outbursts observed between 1926 and 2012. Because the Draconids were widely analyzed by multiple observers using different instruments, the literature has a wide variety of information relating to the shower. For each event, we tried to highlight the conclusions shared by the highest number of authors or the best documented reports.
 
 \item  Sections \ref{sec:model1}, \ref{sec:model1_interp} and \ref{sec:model1_calib}, together with Appendices A, B \& C, describe in detail the implementation and interpretation of the new meteoroid stream model used here to reproduce or predict past and future Draconid occurrences (Model I). In section \ref{sec:MSFC}, we offer a succinct summary of the independently developed NASA Meteoroid Environment Office's MSFC meteoroid stream model, used to validate Model I. 
 
 \item  Section \ref{sec:post_predictions} presents the performance of Model I in reproducing the main visual and radio outbursts observed between 1933 and 2012, in terms of years of appearance and shower time, strength, and duration. A comparison between our simulations and observations of the previously unreported 1999 radio outburst, revealed by our model, is also presented. An initial comparison between the 2018 prediction performed in \cite{Egal2018} and preliminary visual and radio observations of this recent outburst, shared by the International Meteor Organization (IMO), is given in section \ref{sec:2018}. The simulated activity profiles presented in these sections form the main validation of the approach employed by Model I.

 \item Finally, Section \ref{sec:future_outbursts} details our predictions for three potential radio outbursts expected in the next decade (2019, 2025 and 2029). The reliability of this forecast, intrinsically correlated to Model I's accuracy, is discussed in Section \ref{sec:discussion}.  
\end{itemize}

  \section{Draconid observations}\label{sec:observations}
  
 The Draconids are known to have produced two storms observed optically in 1933 and 1946, as well as more moderate shower outbursts in 1926, 1998, and 2011. Other Draconid outbursts were detected mainly by radar techniques in 1952, 1985, and 2005, in addition to a radio storm in 2012. Very low activity from the shower was reported in 1972 instead of the intense outburst/storm that was originally predicted \citep{McIntosh1972,Hughes1973}. 
 
 In this work, we classify the Draconid apparitions into three categories:  the outbursts which were visually observed or recorded using optical instruments (``visual outbursts''), the showers mainly observed by radio/radar devices (``radio outbursts''), and weak or poorly documented showers.  
  
  \subsection{Established visual outbursts}
  
  \subsubsection*{1933}
  The 1933 storm occurred on October 9 around 20h15 UT \citep{Watson1934}, for a total duration of about 4h30 \citep[R. Forbes-Bentley in ][]{Olivier1946}. ZHR estimates of the shower are very disparate and vary from around 5400 \citep{Watson1934} to 10 000 \citep{Jenniskens1995} and even 30\,000 \citep{Olivier1946,Cook1973}.
  
  \subsubsection*{1946}
  The 1946 storm was by visual, photographic, and radar techniques \zcorr{mainly in Europe and North America. The shower peaked around 3h40-3h50 UT on October 10, and lasted 3 to 4 hours \citep{Lovell1947,Kresak1975}. } Again, ZHR estimates in the literature vary between 2000 \citep{Hutcherson1946} and 6800 \citep{Kresak1975} or 10 000 \citep{Jenniskens1995}.  
    
   \subsubsection*{1998}
   The 1998 outburst appeared on October 8, peaking around 13h10 UT \citep{Koseki1998,Arlt1998,Watanabe1999} although \cite{Simek1999} indicated a maximum around 13h35 UT. The total duration was about 4h \citep{Watanabe1999}, and the maximum ZHR reached 700 to 1000 meteors per hour \citep{Koseki1998,Arlt1998,Watanabe1999}.
    
   \subsubsection*{2011} 
   Following the concordant predictions of an outburst in 2011, the Draconids were observed on October 8 by many teams with many different observational techniques (radar, video, photography, visual). The main peak occurred around 20h-20h15 UT, for a total duration of about 3 to 4 hours and a maximum ZHR estimate varying from 300-400 to 560 \citep{Toth2012,Kero2012,Koten2014,Molau2014,McBeath2012,Trigo2013,Kac2015}. 
    
  \subsection{Established radio outbursts}
    
   \subsubsection*{1985}
   The 1985 peak occurred on October 8 between 9h25 and 9h50 \citep{Chebotarev1987,Simek1994}, probably around 9h35 UT \citep{Lindblad1987,Sidorov1994}. The observed shower duration varies from 3h \citep{Lindblad1987,Mason1986} to 4h30 \citep{Sidorov1994}. An equivalent ZHR is estimated to be from around 400-500 \citep{Lindblad1987,Simek1994} to possibly as high as 2200 \citep{Mason1986}. \zcorr{In agreement with the radio measurements, visual observations carried out in Japan in 1985 confirmed a ZHR higher than 500 shortly before 10h UT on October 8 \citep{Koseki1990}.}
     
   \subsubsection*{2005}
   The unexpected 2005 radio outburst was detected on October 8 around 16h05 UT  \citep{Campbell-Brown2006}. The end of the shower was missed by radar, while the beginning was not observed using visual techniques \citep{Koten2007}; producing only a lower limit of 3h for the shower duration. The equivalent ZHR was estimated to be around 150 \citep{Campbell-Brown2006}.
     
   \subsubsection*{2012}
   The 2012 storm happened on October 8 at 16h40, for a peak duration of about 2h and an equivalent ZHR of 9000 \citep{Ye2014}. \cite{Fujiwara2016} measured a total duration of the shower of about 3 hours, with a peak time occurring between 16h20 and 17h40 UT.
  
  \subsection{Weak or controversial shower returns}
  
   \subsubsection*{1926}
   The meteor activity observed by visual observers in October 1926 allowed the shower to be linked to comet 21P/Giacobini-Zinner. From 36 meteors observed between 20h20 and 23h20 (G.M.T) on October 9 1926, a ZHR of about 20 meteors per hour was estimated \citep{Denning1927}. A Draconid fireball, observed around 22h16 (G.M.T) that night, impressed the observers by leaving a persisting train during half an hour \citep{Denning1927,Fisher1934}.
  
  \subsubsection*{1952/1953}
  Draconid activity was noticed by the Jodrell Bank radar on October 9 1952, with a maximum ZHR of 170-180 around 15h 40-50 min UT \citep{Davies1955}. The shower duration was approximately 3 hours. The apparition of Draconid meteors the following year, in 1953, is controversial;  no enhanced activity was detected by the Jodrell Bank radar at any time within 12 hours of the expected maximum time \citep{Jodrell1953}, while other authors suggest some weak activity \citep{Jenniskens2006}.
  
   \subsubsection*{1972}
   Due to the favorable geometrical configuration between the comet and the Earth,  a potentially strong outburst/storm was expected on October 8 1972. However no significant activity was observed \citep{Millman1973}. Radar observations revealed a diffuse component of the shower, with a weak maximum activity at October 8.2$\pm$0.3 \citep{Hughes1973}. Radio observations carried in Japan might however argue for a stronger radio activity, with a peak of 84 meteors detected in a 10 minute interval around 16h10 UT on October 8 1972 \citep{IAU1972}.  \\
   
   Having summarized the major observational features of the Draconids over the last century, our next goal is to reproduce the years of recorded activity, shower timing, strength and activity profile from a model. We attempt to do this employing a model with the fewest number of tunable parameters which also provides good fits to the most robustly determined characteristics of the shower.
  
\section{Model I - Simulations} \label{sec:model1}

\subsection{21P/Giacobini-Zinner}\label{subsec:21P}

 Comet 21P/Giacobini-Zinner was observed for the first time by M.~Giacobini in 1900, and identified again by E.~Zinner in 1913. Since its discovery, at least 16 apparitions of the comet have been observed and numerous orbital solutions produced. 21P is a typical Jupiter-family comet (JFC), with a period of approximately 6.5 years and a current perihelion distance of 1.03 au. The comet has suffered multiple close encounters with Jupiter through its history, resulting \zcorr{in our simulations in a global reduction of its semi-major axis and increase of its eccentricity in 400 years}. The comet's motion has also been affected by sudden and significant variations in the nongravitational forces (NGF) induced by its outgassing, causing 21P to be classified as an ``erratic'' comet \citep{Sekanina1993}. A particular discontinuity in the transverse NGF coefficient between 1959 and 1965 dramatically modified 21P's orbit \citep{Yeomans1989}. The gas and dust production of the comet used to peak after perihelion before this epoch, while the maximum outgassing has occurred pre-perihelion since then \citep{Sekanina1985,Blaauw2014}.

 \subsubsection{Ephemeris}
 
 Several studies have tried to reproduce the comet's dynamical evolution during the 20$^{th}$ century, in particular a significant discontinuity in its orbital evolution observed between 1959-1965. Despite a moderately close approach with Jupiter in 1958, the gravitational influence of the giant planet did not seem to be responsible for the orbital variations observed around 1959 \citep{Yeomans1971,Yeomans1972}. 
 
 \cite{Sekanina1985,Krolikowska2001} were able to explain the NGF evolution of 21P by considering the precession of the comet's spin axis. However, Sekanina's model yielded unrealistic values of the comet's oblateness and rotation period \citep{Yeomans1986,Krolikowska2001}, while the Kr\'{o}likowska analysis required many additional parameters derived from observations (e.g. relative times between the maximum activity and the perihelion passage for different apparitions) to correctly link all the comet's apparitions. Another independent explanation for the erratic orbital evolution of 21P could be the activation of discrete source regions at the surface of the comet, at specific locations and for a certain duration \citep{Sekanina1993}.
 
 In their current state, neither of these models permits the comet's ephemeris to be reproduced without involving a detailed knowledge of its past apparitions; for our study, we therefore decided to rely directly on the observations. The comet's motion at each apparition is integrated from an orbital solution provided by the JPL Small Body Data Center\footnote{https://ssd.jpl.nasa.gov/sbdb.cgi}, with an external time step of 1 day and considering the asymmetric non-gravitational forces model of \cite{Yeomans1989}.
 
 \subsubsection{Integration period}
 
 Because of the 1959 orbital evolution discontinuity, we chose to build the ephemeris of each apparition of the comet from the closest available orbital solution, and not from the most recent measured orbit of 21P.  However, older observations of 21P (prior to 1966) were determined with less accuracy, which may reduce the reliability of our integrations. In order to evaluate the impact of this approach on the comet ephemeris, we compared the trajectories integrated from 15 distinct orbital solutions of 21P obtained for the years 1900, 1913, 1926, 1933, 1940, 1946, 1959, 1966, 1972, 1979, 1985, 1992, 1998, 2006, and 2013.
 
\begin{figure*}[ht]
   \centering
   \includegraphics[width=\textwidth]{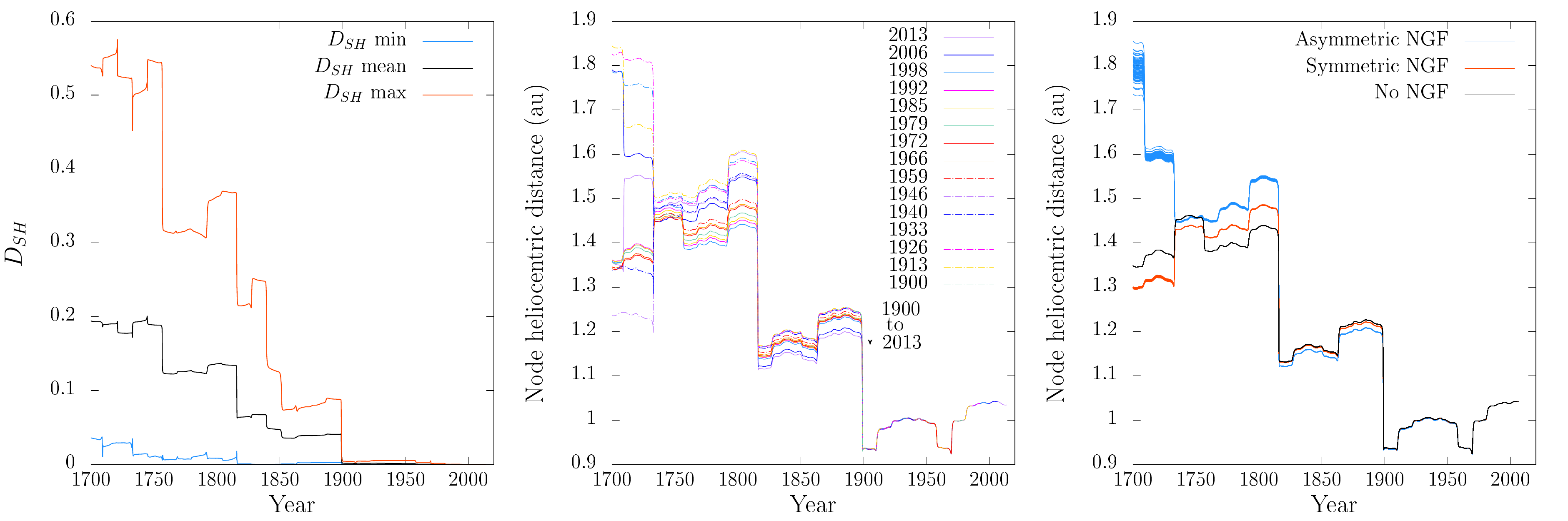}
   \caption{\label{fig:21P_ephem} Left panel: time evolution of the minimum, maximum, and mean $D_{SH}$ between each pair of orbits determined from 15 distinct initial orbital solutions. Middle panel: heliocentric distance of the descending nodes of 21P along time, for 15 initial orbital solutions ranging from 1900 to 2013. Right panel: heliocentric distance of the descending node for different nongravitational forces models.}
   \end{figure*}

 Figure \ref{fig:21P_ephem} illustrates the evolution with time of the minimum, maximum, and average similarity criterion $D_{SH}$ \citep{Southworth1963} for each pair of trajectories obtained (left panel). The middle panel presents the variations of the descending node's heliocentric distance for each initial orbital solution considered. The right panel was obtained by considering the same initial orbit (2006 solution) but integrating the trajectory with different nongravitational force models (no NGF, symmetric outgassing, and asymmetric outgassing, respectively). 
 
 In Figure \ref{fig:21P_ephem}, we observe a slight increase of the $D_{SH}$ around 1960. The $D_{SH}$ evolution before this date coincides with the dispersion noticeable in the node distances plots. The first significant dispersion of the orbits occurs in 1898, as a consequence of a close encounter with Jupiter. Earlier close approaches with Jupiter in 1815 and 1732 also disperse the orbits. From this analysis, we conclude that the orbital solution selected for the comet ephemeris computation should not influence the prediction of the meteor showers caused by trails ejected after the comet's discovery. For older streams, however, the initial integration epoch has a significant impact on the comet ephemeris, and hence on the meteoroid stream generated. Since the Draconids are often associated with young material \citep{Lindblad1987,Wu1995,Vaubaillon2011}, our integrations were limited to the 1850-2030 period. \zcorr{If any interpretation of the contribution of trails ejected prior to 1898 needs to be carefully conducted, their influence on our predictions is negligible. Particles ejected over the 1850-1898 period were only simulated for comparison of the nodal footprints of the simulated Draconids with previous works \cite[e.g. with][]{Vaubaillon2011}.}
 
 
 \subsubsection{Physical properties} 
 
 The lack of direct imaging of 21P's nucleus forces us to estimate certain physical properties needed for our integrations. 
 The radius of the nucleus of 21P is not well determined; estimates suggest it probably lies between 1 and 2 km \citep{Leibowitz1986,Singh1997,Lamy2004,Pittichova2008}. 
 By default, we consider a nucleus density of 400 kg~m$^{-3}$ and an albedo of 0.05 commonly assumed for JFCs \citep{Newburn1985,Landaberry1991,Hanner1992}. The percentage of active surface, unknown for 21P, is fixed to 20\%. A summary of the model's parameters is presented in Table \ref{table:parameters}.

\subsection{Meteoroid streams}

\subsubsection{Ejection}

 Particles are ejected at each simulation time step of the comet for heliocentric distances below 3.7 au \citep{Pittichova2008}. \zcorr{Around 12.48 } million dust particles were simulated over the period 1850-2020, covering the size bins $[10^{-4},10^{-3}]$ m ($10^{-9},10^{-6}$) kg: (\zcorr{160 000 } particles \zcorr{per apparition}), $[10^{-3},10^{-2}]$ m ($10^{-6},10^{-3}$) kg: (\zcorr{190 000 } particles) and $[10^{-2},10^{-1}]$ m ($10^{-3},1$) kg: (\zcorr{130 000 } particles).
 \zcorr{An additional sample of 120 000 particles in each size bin was ejected at each comet apparition between 1852 and 2025, and integrated until the year 2030 (9.72 million particles). In total, 22.2 million dust particles were simulated over the period 1850-2030. }
 
 The meteoroid density is assumed to be 300 kg~m$^{-3}$, as determined by Draconid meteor observations \citep{Borovicka2007}. Simulated particles are isotropically ejected from the sunlit hemisphere of the comet, with velocities determined by the \cite{Crifo1997} model, which produces ejection speeds most in accord with recent \emph{in situ} cometary measurements (c.f. \ref{sec:Vej}).

\subsubsection{Integration} \label{sec:integration}

 The stream integration is performed in Fortran 90 using a 15$^\text{th}$ order RADAU integrator with a precision control parameter LL of 12 \citep{Everhart1985}. The external integration step is fixed to one day and the internal time steps are variable, allowing close encounters with planets to be dealt accurately. The gravitational attraction of the Sun, the Moon, and the eight planets of the solar system as well as general relativistic corrections are taken into account. Solar radiation pressure and Poynting-Robertson drag are also included. The Yarkovsky-Radzievskii effect was neglected since the size of our simulated particles doesn't exceed 10 cm \citep{Vokrouhlicky2000}.   
 
 \subsubsection{Impact selection}
  
  As a first selection, all particles that approach the Earth below a distance $\Delta X = V_r \Delta T$ are considered potential impactors, with $V_r$ the relative velocity between the planet and the particle and $\Delta T$ a time parameter depending on the shower duration. The length of a typical Draconid shower doesn't exceed a few hours; we consider for this study a conservative time interval of $\Delta T = 1$ day and a distance threshold $\Delta X$ of $1.15\times10^{-2}$ au. Potential impacting particles are, in a second step, individually integrated with a time step of one minute, in order to precisely estimate the date and position of their closest approach with the Earth.

   \begin{table*}[!ht]
      \centering
     \begin{tabular}{lcc}
      Comet parameter & Choice & Reference \\
         \hline
         \hline
       Shape & Spherical & \\
       Diameter & 2 km & \cite{Lamy2004}\\
       Density & 400 kg~m$^{-3}$ & \\
       Albedo & 0.05 &  \\
       Active surface & 20 $\%$ & \\
       Active below & 3.7 au & \cite{Pittichova2008} \\
       Variation index $\gamma_1$, $\gamma_2$ & 2.1, 7.3 & This work \\
        $K_1$ & $\sim$ 100 & This work \\
       $Af\rho_{max}$ & $\sim$1600-1800 cm &  This work\\   
           \hline
               & & \\
                  & & \\         
      Particle parameter & Choice & Reference \\
           \hline
           \hline
       Shape & Spherical &  \\
       Size & $10^{-4}$ to $10^{-1}$ m & \\
       Density & 300 kg~m$^{-3}$ & \cite{Borovicka2007} \\
       Ejection velocity model & ``CNC'' model & \cite{Crifo1997} \\ 
       Ejection rate & each day & \\
       N$_\text{particles}$ / apparition &  $\sim$ \zcorr{840 000} & \\
         \hline
     \end{tabular}
     \caption{\label{table:parameters} Comet and meteoroid characteristics considered by Model I. See the text for more details.}
   \end{table*}

 \section{Model I - Interpretation } \label{sec:model1_interp}
 
 The primary goal of this work is to compare our simulated meteor showers to Draconid observations with the goal of matching the peak date, peak intensity, duration and overall activity profiles of past showers. Because of limited computational resources, the number of particles ejected in our model is not comparable to the true number of meteoroids effectively released by comet 21P/Giacobini-Zinner. It is therefore necessary to extrapolate from the comparatively finite number of simulated particles to the true number of particles in the stream.  
 
 Since the number of simulated particles that would physically impact the Earth is too small to derive a reliable flux profile, we consider each particle that passes within a sphere $\mathcal{S}$, centered on the Earth and with a given radius $R_s$, as contributing to the shower. The $R_s$ value must be smaller than the previous $\Delta X$ spatial criterion to exclude surrounding streams not producing any activity at the Earth. On the other hand, $R_s$ needs to be large enough to include a statistically useful sample of particles. The selection of the $R_s$ parameters for the Draconids simulations is discussed in Section \ref{sec:flux}. 
 
 Once particles are selected using the reduced distance criterion $R_s$, we assign to each impactor a weight $N_g$ which represents the scaling required to move from the small number of simulated meteoroids to the true number which would be released by the comet in the same ejection circumstances.

\subsection{Weights}

The weighting process implemented here relies mainly on \cite{Vaubaillon2005}, with some differences. First, the evolution of the dust production for 21P/Giacobini-Zinner with the heliocentric distance $r_h$ is extrapolated to each apparition from a photometric solution of the comet (cf. \ref{sec:comet_photometry}). Second, the gas and dust production are not assumed to be proportional to each other; this decoupling allows the gas and dust production to evolve separately as a function of heliocentric distance \citep{Sekanina1985}. The main hypotheses underlying the weighting solutions are: 

\begin{enumerate}
 \item The nucleus is spherical, homogeneous and composed of dust and water ice.
 \item The dust production of the comet, represented by the $Af\rho$ parameter \citep[][ cf. \ref{sec:comet_photometry}]{AHearn1984}, evolves with the heliocentric distance $r_h$. The gas-to-dust production ratio $K$ is not constant over $r_h$.
 \item The water molecules and the dust particles are ejected from the sunlit hemisphere of the nucleus, with an intensity varying with the angle $\theta$ from the subsolar point.
 \item The size distribution of the particles follows a power law of index $u$.
\end{enumerate}

Details of the weighting computation are provided in \ref{sec:weights_eq}. 
For a given apparition of the comet with a perihelion distance $q$, the number of particles ejected in all directions in the size bin $[a_1',a_2']$ during time $\Delta t$ and around $r_h$ is found to be: 

\begin{equation}\label{eq:weights}
 \left\{
 \begin{aligned}
 N_g=& K_2\frac{J\Delta t}{2 A(\Phi)}A_1(a_1',a_2')\zcorr{\cdot}\left(K_1\Delta r_h\color{white}\frac{1}{2}\color{black}\right.\\
 -&\left.\frac{\zcorr{Af\rho_{max}}}{\gamma \ln(10)} \left[10^{-\gamma (r_h+\frac{\Delta r_h}{2}-q)}-10^{-\gamma (r_h-\frac{\Delta r_h}{2}-q)}\right] \right)\\
 \gamma=&\gamma_1 \text{ pre-perihelion and } \gamma=\gamma_2 \text{ post-perihelion}
 \end{aligned}
  \right.
 \end{equation}
where:
  \begin{itemize}
  \item[-] $K_2$ is a tunable normalization coefficient constant over all the ejection conditions,
  \item[-] $A(\Phi)$ is the nucleus albedo at phase angle $\Phi$,
  \item[-] $Af\rho_{max}$ is the maximum $Af\rho$ obtained for a given apparition (ejection epoch),
  \item[-] $r_h$ is the meteoroid heliocentric distance at the time of ejection (in au),
  \item[-] $q$ is the perihelion distance (au),
  \item[-] $A_1(a'_1,a'_2)$ is shorthand for $\int_{a'_1}^{a'_2}\frac{da}{a^{u}}$, with $a$ the particle radius (in m),
  \item[-] $K_1$ is a constant governing the $Af\rho=f(r_h)$ evolution (fit parameter),
  \item[-] $\gamma$ is the $Af\rho$ variation index, with $\gamma=\gamma_1$ for pre-perihelion locations and $\gamma=\gamma_2$ for post-perihelion measurements, and
  \item[-] $J$ is a function of the ejection velocity and the minimal and maximal particle radii that can be ejected from the comet (cf. \ref{sec:weights_eq}).
  \end{itemize}
  
  \subsection{Flux and activity profile} \label{sec:flux}

  The shower spatial number density $\mathcal{F}_s$ is estimated from the weighted number of particles entering the sphere $\mathcal{S}$ over time. In this work, the sphere radius is fixed to $R_s=V_\oplus \delta t$, with $V_\oplus$ the Earth's \zcorr{heliocentric } velocity and $\delta t$ a time parameter corresponding to the duration of the maximum activity of the shower. For the most intense Draconid outbursts (e.g. 1946), most of the activity was contained within a one-hour interval \citep{Davies1955,Kresak1975}. For the main Draconid outbursts observed visually (1933, 1946, 1998 and 2011), we then choose $R_s=V_\oplus \times 1$ $h$ ($R_s < 10$ Earth radii). When the total number of unweighted particles retained for the flux computation is lower than 15, $\delta t$ is gradually increased until either the number of particles exceeds 15 or $\delta t$ is 6 hours (assumed to be the maximal duration of the shower).  The flux $\mathcal{F}$ of meteoroids is obtained by:
  
  \begin{equation}
   \mathcal{F}=\mathcal{F}_s\zcorr{\cdot V_r}
  \end{equation}
  
  \noindent with \zcorr{$V_r$ } the relative velocity between the Earth and the meteoroid stream. The zenithal hourly rate  of the shower is derived from $\mathcal{F}$ using \cite{Koschack1990}: 
  
  \begin{equation}\label{eq:ZHR}
  \textrm{ZHR}=\frac{\mathcal{A_\text{s} F}}{(13.1r-16.5)(r-1.3)^{0.748}}
  \end{equation}
  
  \noindent where $r$ is the measured population index of the shower and $\mathcal{A_\text{s}}$ \zcorr{an observer's } surface area of the atmosphere at the ablation altitude ($\mathcal{A_\text{s}}\sim 37200\; \mathrm{km}^2$).
  
  \subsection{Peak time estimate} \label{subsec:peak_time}
  
  The predicted peak date of the shower is determined by computing a weighted average of the simulated activity profile. The presence of additional peaks, as well as the intensity and the duration of the shower, are also derived from the ZHR evolution. In the case of simulated outbursts with a low number of particles retained for the flux computation ($\delta t > 1 h$), the resolution of the activity profiles may compromise the validity of this approach. In such cases, the time of the peak maximum is taken to be the date of closest approach between the Earth and the median location of the meteoroid streams \citep{Vaubaillon2005}.
  
  \section{Model I - Calibration} \label{sec:model1_calib}
 Before predicting any future Draconid showers, we need to determine the unknown parameters ($K_2,u$) of the weighting function in Eq. \ref{eq:weights} which best reproduce the past observed outbursts. The results are particularly sensitive to the value of the size distribution index $u$ of the meteoroids at ejection, since this value influences the intensity and the shape of the activity profiles as well as the peak time estimate. 
 
 The calibration is first performed on the four strong visual showers of 1933, 1946, 1998, and 2011 and then refined using the radio outbursts. The simulated ZHR profiles, derived from the meteoroid fluxes using Equation \ref{eq:ZHR} and a given population index $r$, are compared with the observations \zcorr{(see Figure \ref{fig:ZHR_profiles} later)}. We have chosen in the calibration process not to allow a variable population index, but instead adopt a fixed value of $r=2.6$ based on measurements of the 2011 \citep{Toth2012,Kac2015} and potentially the 1933 shower \citep{Plavec1957}. 
 
 Our philosophy is that since any model can be improved when more free parameters are introduced, we chose to limit the number of tunable parameters to the absolute minimum required. In doing this, we decrease the match between the simulated and observed intensities of a specific shower in fixing $r$. However, this allows us to determine the weighting solution that best reproduces all the outbursts with minimum adjustable parameters, which we feel improves the robustness of the model shower prediction. 
 
 The best agreement between the simulated and the observed activity profiles for the four visual showers was obtained for an ejecta size distribution index $u$ of 2.9. This value differs from our previous published estimate of 2.64 \citep{Egal2018}; differences between the two works are due to several factors. First, in this work, we have increased the number of simulated particles by 75\% and used a slightly different weighting solution which better reflects the observed $Af\rho$ evolution of comet 21P. Second, the size distribution index used here was determined considering two additional outbursts (1999 and 2018), not included in \citep{Egal2018}. 
 
 Our calibrated value of $u$ lies in the range
 of measurements performed for 67P \citep[drifting from 2 beyond 2 au to 3.7 at perihelion, cf.][]{Fulle2016}. The associated cumulative mass index ($s_{mc}=0.63$ for $u=2.9$) is somewhat lower than the value of 0.85 derived for comets 1P/Halley \citep{McDonnell1987} and 81P/Wild 2 \citep{Green2004}.  
 Unfortunately, no such \emph{in situ} measurement have been performed for 21P. Indirect estimates of $u$, inferred by meteor observations, are hampered by the variability of the observed mass index changes with each apparition of the shower and even during an individual outburst \citep{Koten2014}. In addition, because the orbital evolution of a meteoroid is size-dependent, a fundamental difference between the size distribution index at the time of ejection and observation is not unexpected. Since no \emph{in situ} or indirect measurement of this parameter has been performed for 21P, the value of $u=2.9$ will be adopted through the rest of this study.

  \section{Model II - MSFC}\label{sec:MSFC}
  
   For validation purposes, the Draconid streams simulated using Model I have been compared to an independent meteoroid modeling performed by the NASA Meteoroid Environment Office (MEO). The MEO's MSFC Meteoroid Stream Model, similar to the model presented above, is detailed in \cite{Moser2004} and \cite{Moser2008}. The position and velocity of 21P/Giacobini-Zinner is taken from the 2006 orbital solution and ephemeris provided by JPL HORIZONS\footnote{https://ssd.jpl.nasa.gov/horizons.cgi}. The comet radius is assumed to be 1.7 km, i.e. an average of \cite{Landaberry1991,Churyumov1991} and \cite{Newburn1989} measurements. 600 000 meteoroids are ejected at each apparition of the comet between 1594 and 2018, resulting in a total of about 35.4 million particles. Particles are released from the comet with a time step of 1 hour, using the ejection model of \cite{Jones1995} with a spherical cap angle of ejection of 30$\degree$. The size of the particles are distributed uniformly over $\log\beta$, where $\beta$ ranges from $10^{-5}$ to $10^{-2}$. $\beta$ is defined as the ratio between the forces due to the radiation pressure ($F_{rad}$) and the Sun's gravity ($F_{grav}$), and can be approximated by the following relation: 
   
   \begin{equation}
       \beta=\frac{F_{rad}}{F_{grav}}\simeq \frac{5.7\times10^{-4}}{a \rho}
   \end{equation}
   
   with $a$ and $\rho$ respectively the radius (m) and density (kg~m$^{-3}$) of the particle. The selected $\beta$ range of $10^{-5}$ to $10^{-2}$ corresponds to masses between 1 $\mu$g and 1 kg for the selected density of 1000 kg~m$^{-3}$, and masses around ten times higher for the density of 300 kg~m$^{-3}$ considered by Model I. 
   
   The meteoroid integration is performed exactly as described in Section \ref{sec:integration}. Particles are retained as potential impactors if they cross the Earth's orbital plane within 0.01 au of the planet, and within $\pm$ 7 days of the expected shower peak. The time and intensity of the shower's maximum activity are evaluated through computation of an impact parameter (IP), defined as 
   \begin{equation}
   \textrm{IP}=\frac{R_\oplus + h_{atmos}}{D}
   \end{equation}
   \noindent where $R_\oplus$ is the Earth's radius, $h_{atmos}$ the height of the atmosphere, and $D$ the Earth-particle distance at the time of nodal crossing. The impact parameter increases the weight of particles passing by the Earth more closely, and therefore more susceptible to contribute to a shower.  A Lorentzian fit of the IP distribution as a function of the solar longitude allows the time of the shower peak to be estimated. The shower strength is usually estimated by scaling the IP distributions to historical observations.
   
   Since less time and effort was allocated for the analysis of the MSFC simulations, the output of Model I and Model II cannot be compared for all the shower's characteristics. Results of the MSFC model, successful in predicting previous Draconid outbursts (e.g. in 2011), are however presented in this work because they allow an assessment of the confidence level of the time and intensity predictions issued from Model I.

  \begin{table*}[!ht]
   \centering
     \begin{tabular}{lcc}
  Comet parameter & Choice & Reference \\
        \hline
     \hline
     Shape & Spherical & \\
     Radius & 1.7 km & See text \\
     Active within & 2.5 au &  \\ 
     \hline
     & & \\
     & & \\         
     Particle parameter & Choice & Reference \\
          \hline
     \hline
     Shape & Spherical &  \\
     Size ($\beta$) & $\sim 10^{-5}$ to $10^{-2}$ & \\
     Density & 1000 kg~m$^{-3}$ &  \\
     Ejection velocity model &  & \cite{Jones1995} \\ Cap angle & 30$\degree$ & \\
     Ejection rate & every hour & \\
     N$_\text{particles}$ / apparition &  $\sim$ 600 000 & \\
     \hline
     \end{tabular}
     \caption{\label{table:parameters2} Comet and meteoroid characteristics used by the MSFC model (Model II).}
   \end{table*}    

\section{Outbursts post-prediction} \label{sec:post_predictions}
  
  The goal of this section is to investigate the agreement of our simulated activity profiles with historical observations of the Draconids in terms of peak time, peak intensity, and shower duration. In Section \ref{sec:annual_profile}, the years of potential apparitions of the shower are appraised from the simulations. Details of each visual and radio outburst mentioned in Section \ref{sec:observations} are detailed in Section \ref{sec:past_activity_profiles}.
    
  \subsection{Validation} \label{sec:annual_profile}

  \subsubsection{Annual activity profile}
  
\begin{figure*}[!ht]
    \includegraphics[width=\textwidth]{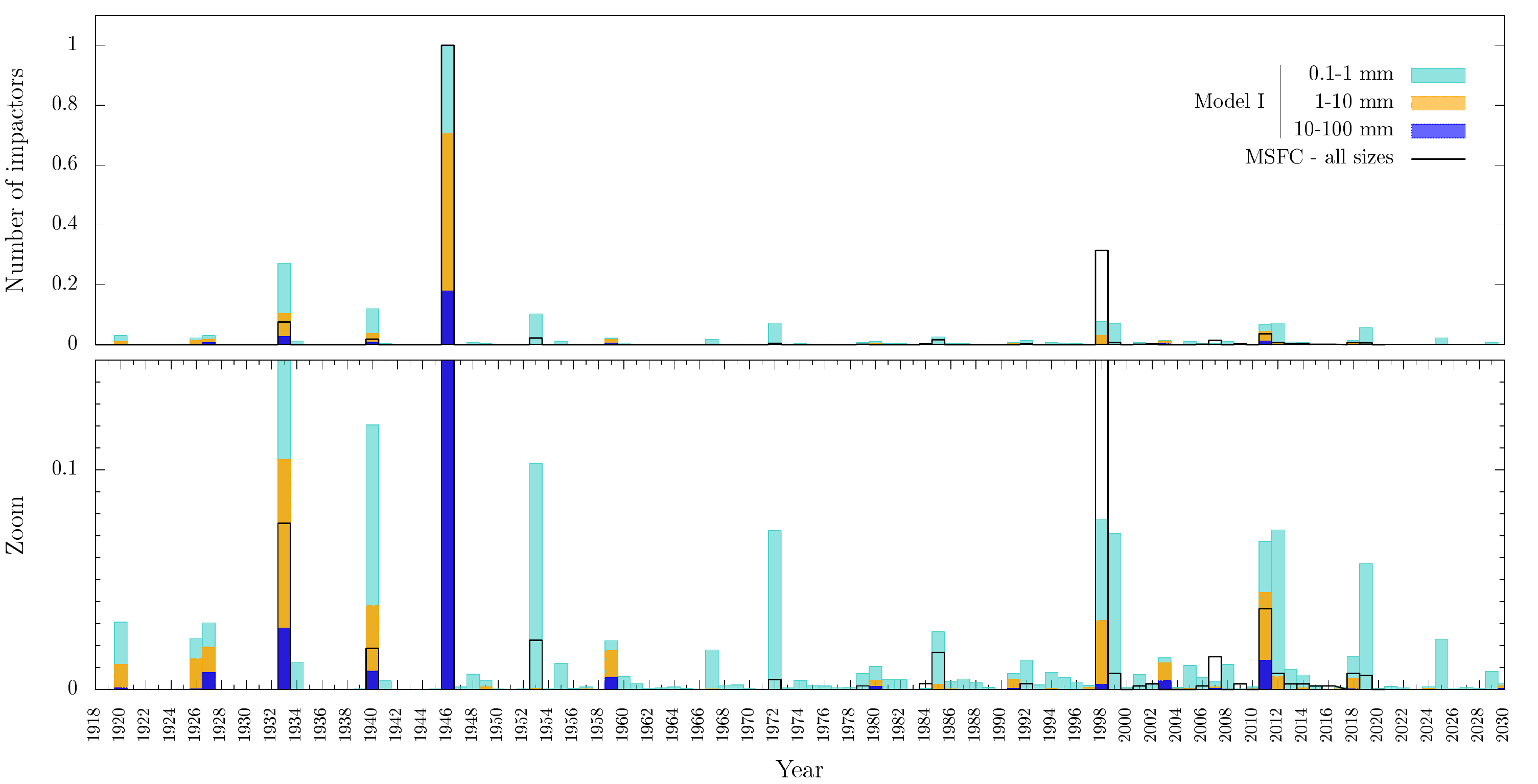}
     \caption{\label{fig:years_activity} The number of impactors (normalized to the number of particles in 1946) reaching the Earth over the period 1920-2030 for each size bin simulated by Model I (filled boxes) and for all the particles generated using Model II (bold line, MSFC model).}
  \end{figure*}

  Years of activity caused by particles of the three size bins simulated by model I are presented in Figure \ref{fig:years_activity}. In this diagram, all the particles meeting the large distance criterion $\Delta X=1.15\times 10^{-2}$ au for Model I and below 0.01 au for Model II are included. The annual variation of the number of impactors presented in Figure \ref{fig:years_activity} is therefore not a good proxy for the shower's strength, but instead highlights the periods of potential activity.

  From our simulations, we predict intense Draconid activity in 1933, 1946, 1998, and 2011 for particles with a size larger than $10^{-3}$ m (i.e., those visible to the naked eye). We also predict activity in 1940, which to our knowledge was not reported by any observer, as well as some in 1959, 2018, and around 1926. When considering smaller particles in the size bin $[10^{-4},10^{-3}]$ m, we see additional  significant activity in 1953, 1972, 1985, 1999, 2012, and 2019, as well as minor contributions over the period 1985-2030 (including the 2005 outburst).

  Both Model I and Model II were successful in reproducing all the years for which a Draconid shower was observed, except the 1952 radio outburst detected only by the Jodrell Bank radar \citep{Jodrell1953}. For Model I, all the visual showers are clearly distinguishable in the profile while the faintest radio events are more complicated to identify (e.g. 2005). Figure \ref{fig:years_activity}, shows that the models are good at reproducing the years of potential enhanced Draconid activity.

  \subsubsection{1999 outburst} \label{subsubsec:1999}
    
 Our model predicted strong activity rich in smaller particles on October 9 1999, which would be detectable using radio instruments. This model outcome was robust and  persisted despite several attempts (not detailed here) to change the weighting parameters and distance criteria  to remove this seemingly non-shower outburst year while still allowing fits to other years. 
 
 Few observations of Draconid activity in 1999 are recorded in the literature and no evidence for a strong radio shower was reported in the literature at the time. Some visual observations conducted from Europe on October 8 indicated low-level Draconid activity \citep{Langbroek1999}. Visual observations performed in Japan on October 9 also indicated a weak Draconid shower, with ``a broad peak from 10h00m to 13h00m UT'' and meteor rates small compared with those in 1998 \citep[ZHR = 20-30, cf.~Iiyama in][]{Sato2003}. These observations are consistent with the simulations, which predict intense activity of particles detectable by radar, but too small to be seen optically.

  In 1999, \cite{Brown2000} conducted a Leonid observation campaign in the Canadian Arctic (at Canadian Forces Station Alert, Nunavut) involving an automated meteor radar. The radar, an early version of the Canadian Meteor Orbit radar (CMOR), consisted of the 29 and 38 MHz systems, but lacked the outlying receivers needed for orbital calculations. It was located at 82.455$^\circ$N, 62.497$^\circ$W. Only the 29 MHz system was in operation on October 9, 1999 as part of early calibration prior to the Leonids; its transmitter power was 3.46 kW. Data from this calibration interval had been collected but not examined in detail prior to the current study, at which time a strong outburst from the Draconids was found in the calibration data.
  
\begin{figure}[!ht]
        \centering
         \includegraphics[width=.48\textwidth]{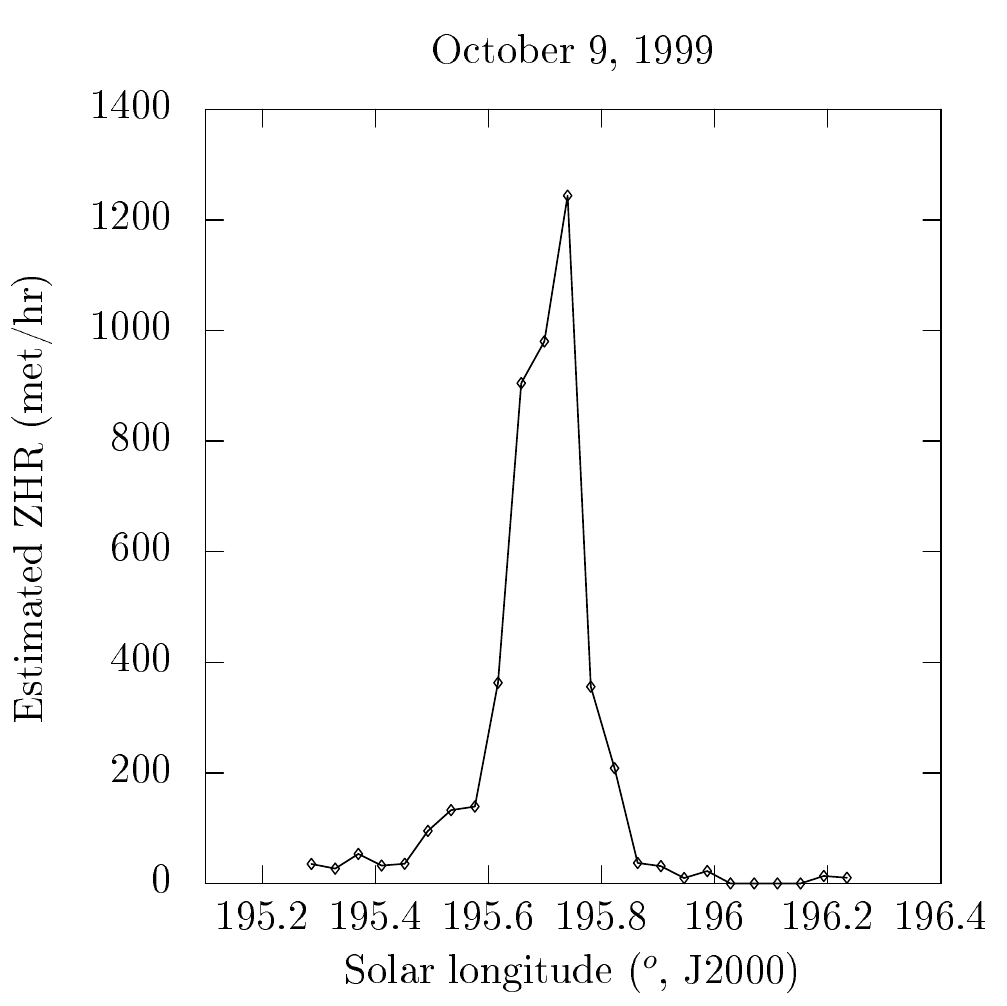}
         \caption{\label{fig:1999} Draconid ZHR profile determined from radar measurements performed at Alert on \zcorr{October 9, 1999}, assuming a mass index of $s=1.82$.  }
        \end{figure}

  The radar echo data were processed to produce fluxes with the same single-station procedure as in \cite{Campbell-Brown2006}, in which echoes occurring at 90$^\circ$ to the radiant are counted and background subtractions are applied. The collecting area of the radar was near maximum at the time of the observed shower, ranging from 500 to 750 km$^2$. The Draconid radiant was close to its lowest elevation of 50$^\circ$. Approximately four hundred echoes occurred on the Draconid echo line in the six hours surrounding the peak, compared to 30 sporadics the previous day. Figure \ref{fig:1999} presents the ZHR profile of the shower with time bins of 15 minutes. The peak activity occurred at 11h30$\pm$15m, for an approximate ZHR of 1250 \citep[assuming a mass index of 1.82, c.f.][]{Pokorny2016}. Unfortunately, no other radar dedicated to meteor observation and recording in October 1999 has been identified.

  Since our simulations allowed us to correctly reproduce the years of noticeable Draconid activity and provided a successful ``prediction'' of the 1999 radio outburst, we consider at this stage Model I to be validated. Next we investigate each simulated shower in more depth. 

\subsection{Activity profiles} \label{sec:past_activity_profiles}
  
\begin{figure*}[!ht]
   \includegraphics[width=\textwidth]{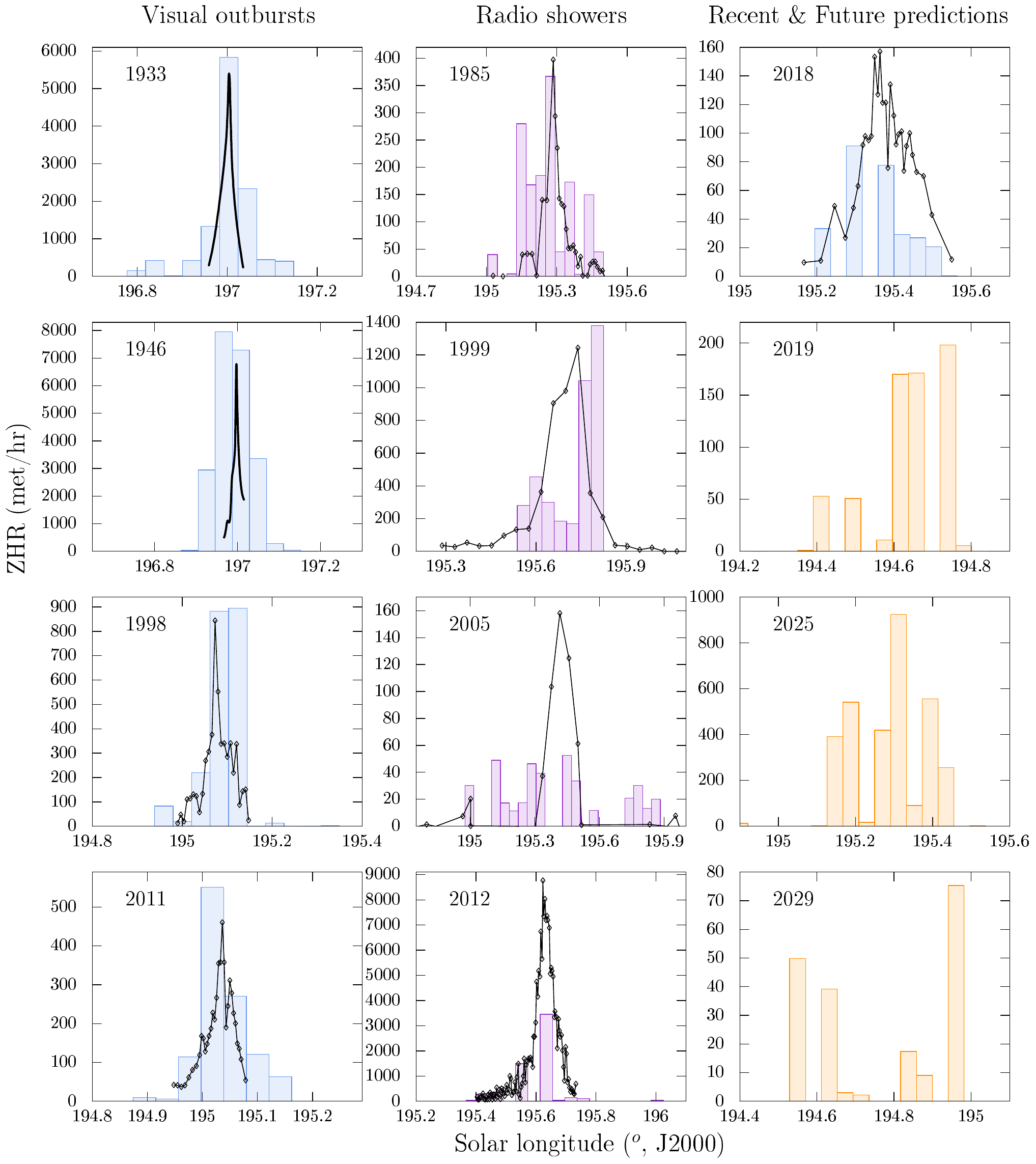}
   \caption{\label{fig:ZHR_profiles} Observed (black solid curve) and simulated (filled boxes) ZHR profiles of nine historic Draconid outbursts between 1933 and 2018 and predictions of three future occurrences of the shower. References for the observations are 1933: \cite{Watson1934}, 1946: \cite{Kresak1975}, 1998: \cite{Watanabe1999}, 2011: \cite{Kac2015}, 1985: \cite{Mason1986}, 2005: \cite{Campbell-Brown2006}, 2012: \cite{Ye2014}. The 1999 profile is compared to radar measurements performed in Alert (cf. Section \ref{subsubsec:1999}) and the 2018 profile is compared with the ZHR derived from the IMO VMDB, accessed 15/12/2018. }
  \end{figure*}

  ZHR profiles derived for each year of activity are presented in Figure \ref{fig:ZHR_profiles}. The first column of the figure groups the four main visual outbursts (1933, 1946, 1998, and 2011) and the second column the radio showers (1985, 1999, 2005, and 2012), whose characteristics are detailed below. The third column displays our simulations of the recent 2018 Draconids and of three potential future outbursts (in 2019, 2025 and 2029). These four cases are detailed in Sections \ref{sec:2018} and \ref{sec:future_outbursts}.
  
  When available, observed profiles (solid lines) have been superimposed on the simulation results (filled boxes). The maximum ZHR (number of meteors per hour) is compared here to the profile's maximum (using a time bin of 1 hour for the model) while the peak time is estimated considering model time bins of 20 minutes. All the time estimates presented in this work have an accuracy of about half an hour. A summary of the main results shown in Figure \ref{fig:ZHR_profiles} is provided in Table \ref{table:results_simulations}. 
  
  \subsubsection{Established visual outbursts}
 
\subsubsection*{1933}
From the modelled activity profile, we predict a maximum on October 9, around 20h13 ($L_\odot=197.005\degree$), reaching a ZHR of 5830. The outburst is caused by particles ejected during the 1900 and 1907 apparitions, with a minor contribution from the 1886 and 1894 perihelion approaches. Most of the particles (71\%) belong to the size bin $[10^{-3},10^{-2}]$ m, with a significant contribution of smaller particles (23\% $\in[10^{-4},10^{-3}]$ m) and some large particles (5\%). The simulated outburst duration is 5 hours, consistent with the observations.  
  
\subsubsection*{1946}
  The simulated maximum occurs on October 10, around 3h38 ($L_\odot=196.986\degree$), reaching a ZHR of 7960. The outburst is caused by particles of several streams ejected between 1900 and 1940, with the largest contributions from the 1900 and 1907 trails. The vast majority of particles belong to the visible range, with a large number of particles having sizes between $10^{-2}$ and $10^{-1}$ m (65\%). The total duration of the simulated outburst is 4h30, in good agreement with measurements. \zcorr{The predominance of the 1900 and 1907 trails influence on the 1933 and 1946 storms is consistent with the results of \cite{Reznikov1993} and \cite{Vaubaillon2011}. }
  
  \subsubsection*{1998}
  The simulated activity peaks on October 8, around 13h26 ($L_\odot=195.087\degree$), reaching a ZHR of about 900. The outburst is caused by particles of the 1926 trail only, mainly with sizes in the ranges $[10^{-3},10^{-2}]$ m (67\%) and $[10^{-4},10^{-3}]$ m (29\%). The estimated shower duration is 4h. Contrary to \cite{Sato2003}, we don't need to assume weak activity of the comet in 1926 to explain the intensity of the 1998 outburst compared to the 1933 and 1946 storms. 
  
  \subsubsection*{2011}
  The modelled maximum activity occurs on October 8, around 20h06 ($L_\odot=195.033\degree$), reaching a ZHR of 550. The outburst is mainly caused by particles from the 1900 and 1907 streams, potentially leading to two maxima around 19h30 and 20h-20h15 (cf. Section \ref{sec:observations}). However, the resolution of the simulated profile of Figure \ref{fig:ZHR_profiles}  does not permit reliable identification of the observed secondary peak. From our simulations, the outburst is mostly composed of particles of size $[10^{-3},10^{-2}]$ m (62\%), with the remainder split evenly between smaller and larger particles. The estimated total duration of the shower is 4h, consistent with observations.

  \subsubsection{Established radio outbursts}
  
  \subsubsection*{1985}
  The simulated maximum occurs on October 8, around 9h42 ($L_\odot=195.260\degree$), reaching a ZHR of 370. The outburst is caused by the stream ejected in 1945, with a minor contribution from the 1933 apparition. Most of the particles contributing to the ZHR belong to the radio range ($\in[10^{-4},10^{-3}]$ m, 89\%) with some candidates between $10^{-3}$ and $10^{-2}$ m (11\%). No larger particles from the simulations were found to encounter the Earth. The total simulated duration (including all the activity detected and not only the central activiy plateau) is slightly shorter than 9h. 
  
  \subsubsection*{1999}
  The simulated activity peaks on October 9, around 11h13 ($L_\odot=195.728\degree$), reaching a ZHR of 1380. The outburst is exclusively caused by small particles ($\in[10^{-4},10^{-3}]$ m, 100\%) ejected in 1959 and 1966. For this simulated profile, we encountered an unusual situation. A small number of particles ($<10$) belonging to the 1966 trail fulfilled all the ejection circumstances required to increase their weight (ejection at perihelion, low phase angle, very small size, etc.). This combination led to weights more than 10 times higher than any other of the 22.7 million particles simulated. Taken at face value this would produce a maximum ZHR of about 50 000 meteors per hour. Given the exceptional nature of the ejection and excessive weighting compared to all other simulated particles, they were removed from the analysis as being unphysical and from the profile of Figure \ref{fig:ZHR_profiles}. The total duration of the simulated outburst is about 6h45. 
  
  \subsubsection*{2005} 
  The 2005 profile represents the worst match between our simulations and the observations. In exploring matches, we consistently found that  weighting parameters adopted which better reproduce this outburst, led to a general model solution which was poorer for all other shower return fits. Allowing this poor fit between the simulations and observations for the 2005 return we feel is justified by the low intensity of the shower and the reduced confidence in the time estimates derived from the observations \citep{Campbell-Brown2006,Koten2007}. 
  
  From the coarse simulated profile, we estimate a maximum activity on October 8, around 16h06 ($L_\odot=195.401\degree$), reaching a ZHR of 50, composed only of particles in the radio range. Contrary to the simulations in  \cite{Campbell-Brown2006}, the best agreement with the observed time and intensity was found for particles ejected in 1953 and not 1946. Differences in the comet ephemeris might explain this divergence. For this poorly simulated profile, a maximum duration of 7h is expected.  

  \subsubsection*{2012}
\begin{figure*}
 \centering 
 \includegraphics[width=.93\textwidth]{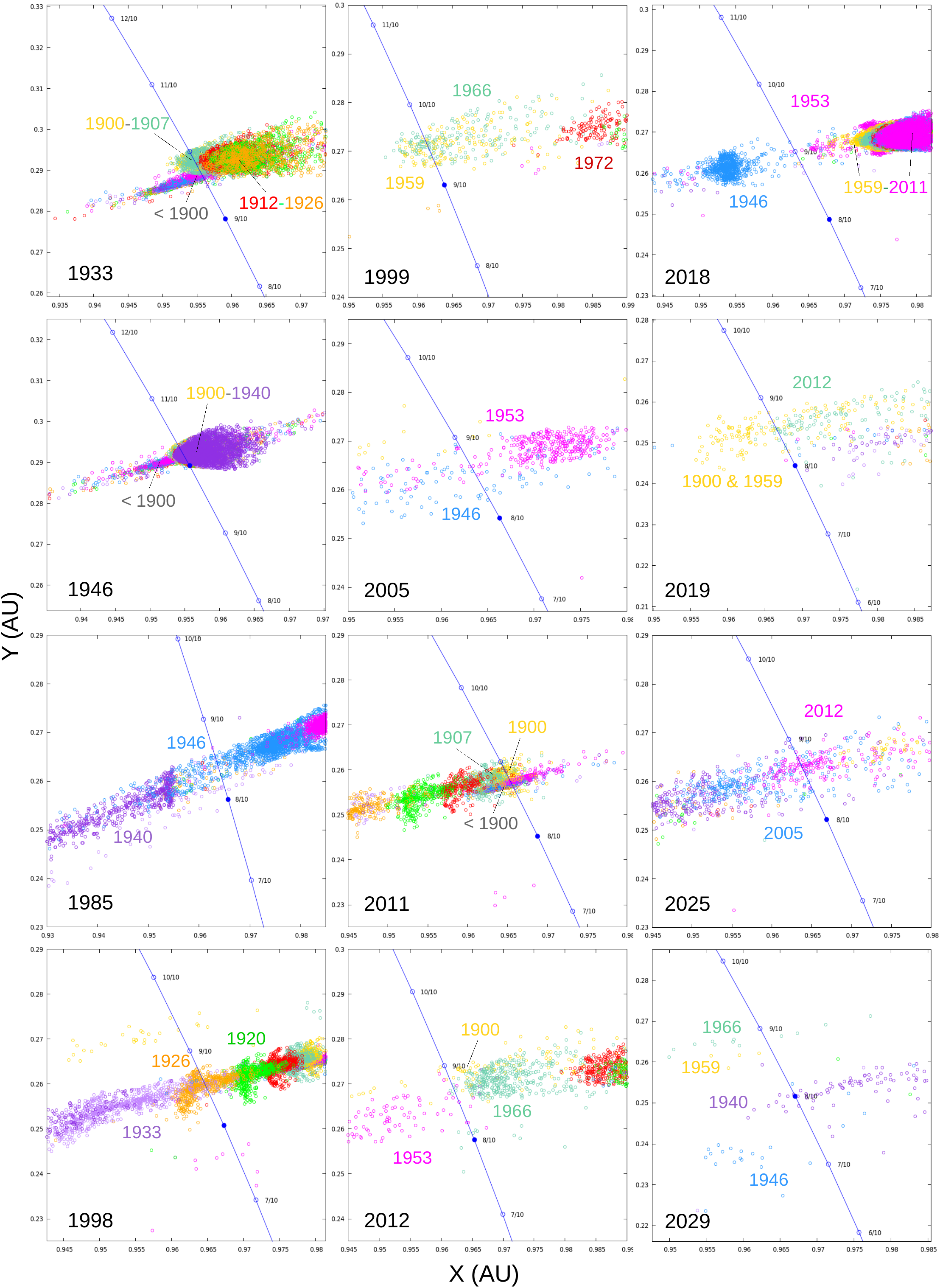}
 \caption{\label{fig:nodes} Nodal footprint of the simulated Draconids crossing the ecliptic plane for years of predicted enhanced activity. The line connects the Earth's positions at the considered dates.}
 \end{figure*}

  The simulated maximum occurs on October 8, around 16h20 ($L_\odot=195.610\degree$), reaching a ZHR of 3450. The outburst is only caused by small particles belonging to the 1966 stream. Similar to \cite{Ye2014}, we find that the nodal footprint of the 1966 trail does not cross the Earth's trajectory (cf. Figure \ref{fig:nodes}). This might account for our underestimate of the shower's intensity; the 2012 outburst reached storm level with about 9000 meteors per hour. However, the particles reaching the planet's vicinity allow us to derive an activity profile that is consistent in time with the observations, with a total duration slightly shorter than 11h. 
  
  \subsubsection{Post-prediction summary}
  
   As a result of the calibration, each simulated profile of a visual outburst accurately matches the observations in terms of peak time ($<$ 30 min) and maximum intensity (within a factor of 2, consistent with the uncertainties related to the meteor observations). The duration of the simulated profiles slightly exceeds the real duration of the shower, but still allows the period of noticeably enhanced activity to be constrained. 
   
   By applying the same weighting parameters to all observed showers, the simulated profiles are less able to reproduce details of the Draconid radio outbursts. Because of the lower number of small model particles reaching the Earth ($N_{part}\simeq 15$), all the radio profiles presented in this section have irregular shapes, with gaps in solar longitude that make their comparison to radar measurements difficult. However, both peak time estimates provided in Table \ref{table:results_simulations} are consistent with the observed maximum activity with an error of less than half an hour. The shower strength, which is much more variable, matched the observations within a factor of 3 in the worst cases such as 2005 and 2012. Despite the apparent discrepancies between the shower duration reported in the literature and our simulations (e.g. in 1985), a quick look at Figure \ref{fig:ZHR_profiles} confirms that we correctly constrained the time window of most of the radio showers. 
   
   For the first time, to our knowledge, we have presented in this section simulated ZHR profiles that allow us to quantitatively estimate the peak time, intensity, and duration of the main Draconid outbursts with reasonable accuracy.

      \begin{table*}[!ht]
        \centering 
        \resizebox{\textwidth}{!}{
        \begin{tabular}{c|ccc||cccc|cc}
         \multicolumn{1}{c}{ } & \multicolumn{3}{c}{ Observations} & \multicolumn{5}{c}{Simulations} & \multicolumn{1}{c}{ }   \\[0.1cm]
         \multicolumn{1}{c}{ } & \multicolumn{3}{c}{ } & \multicolumn{4}{c}{Model I}  & \multicolumn{2}{c}{ MSFC model} \\[0.1cm]
         Date & Time  & Reported & ZHR & Time & Total  & ZHR & Time*  & Time   & ZHR    \\
         \zcorr{(y/m/d)}    & (UT) &  duration        &     & (UT) &   duration        &    & (UT) & (UT)        &        \\
         \hline   
              &     &      &                 &          &     &       &          &        \\  
         \zcorr{1933/10/09}   & 20h15 & 4h30     & 5400 to 30 000 & 20h13     & 5h  &  5830   & 20h08    & 20h14 & strong \\
         \zcorr{1946/10/10}  &  3h40-50 & 3-4h   & 2000 to 10 000 & 3h38     & 4h30 &  7960  &  3h39   & 4h05 & storm \\
         \zcorr{1985/10/08}   & 9h25-50 & 4h30   & 400 to 2200 &  9h42 & 8h30  & 370 &  10h43    & 10h33 & outburst  \\
         \zcorr{1998/10/08}   & 13h10 & 4h       & 700 to 1000 & 13h26 & 4h  & 900 & 13h22 & 13h19 & strong \\
         \zcorr{1999/10/09}   & 11h15-45 & 5h & 1250  & 11h13  &  6h45  & 1380  & 11h30 & 11h57      & moderate     \\
         \zcorr{2005/10/08}   & 16h05 & $\geq$ 3h & 150 & 16h06  & 7h  & 50 & 16h28 & \zcorr{Oct. 9}, 3h52 & weak    \\
         \zcorr{2011/10/08}   & 20h00-15 & 3-4h  & 300-400 to 560 & 20h06 &  4h   & 550 & 20h03   & 19h48 & outburst \\
         \zcorr{2012/10/08}   & 16h40 & 2h  & 9000  & 16h20 & 10h30 &  3450   & 15h58  &   16h42   & weak       \\
         \zcorr{2018/10/08} & 23h15-45 & 3h30$^{a}$  & 100-150 & 22h51 & 8h & 90 & 23h45 & 0h30 & moderate \\
        \end{tabular}}
             \caption{\label{table:results_simulations} Summary of the observed (left panel) and simulated (right panel) characteristics of the historic Draconid outbursts. *Peak time estimated from the stream median location; see Section \ref{subsec:peak_time}. $^{a}$: reported full width at half maximum in 2018 (Molau, personal communication) }
       \end{table*}  
       
 \section{Draconids 2018} \label{sec:2018}
 
  \subsection{Predictions}
  
  From the simulation of 12 million particles using Model I and a slightly different weighting scheme, we performed a prediction of the 2018 Draconids in \cite{Egal2018}. As found in several other studies, our work suggested that the Earth would cross the meteoroid stream through a gap between the 1946 and 1953 trails. The extremely small number of particles reaching the Earth's vicinity and retained for the flux computation in this orbital configuration prevented us from confidently predicting a complete simulated profile. The peak time was therefore estimated from the date of the closest approach to the stream median location (cf. Section \ref{subsec:peak_time}). This approach, which we have found to be more robust in cases with low numbers of model particles, led to a maximum activity estimate around 23h51 ($L_\odot=195.390 \degree$) on October 8, 2018. The model maximum ZHR profile being about 80 (with a factor 3 uncertainty), we had predicted an activity not exceeding a few tens of meteors per hour. However, our predictions at the $L_1$ and $L_2$ Lagrange points suggested storm-level activity at these locations, leading to the temporary re-orientation of the Gaia satellite (personal communication from Serpell, 2018). \\
  
  The updated simulated profile for 2018, calibrated using the larger sample of \zcorr{about 22 } million particles, is presented in Figure \ref{fig:ZHR_profiles}. Compared to our previous work \citep{Egal2018}, we note a slight modification of the profile shape, but no significant variations of the maximum intensity (90 meteors per hour) or time (23h45 using the median location method, $L_\odot=195.348\degree$). In our simulations, all the particles reaching Earth belong to the 1953 trail, and have relatively small sizes ($[10^{-3},10^{-2}]$ m: 87\%, $[10^{-4},10^{-3}]$ m: 12\%,). Almost no large particles encountered the Earth. 
  
   \begin{figure}[!ht]
   \centering
   \includegraphics[width=0.48\textwidth]{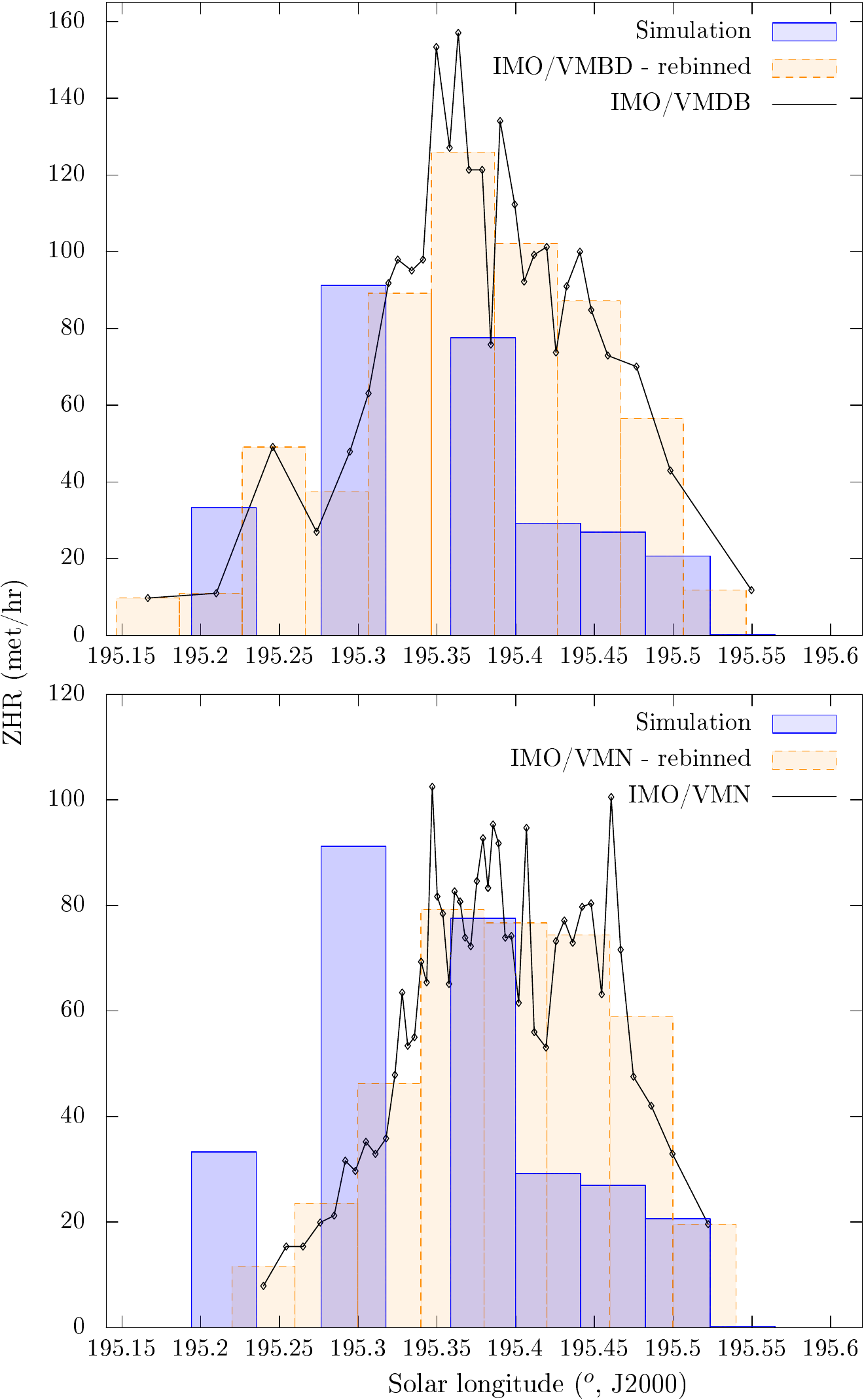}
   \caption{\label{fig:2018} Comparison between the 2018 simulated profile (dark boxes) and preliminary observations released by the International Meteor Organization. Visual observations are shown in the top panel \& video observations in the bottom panel; the simulation results are identical in the two panels.}
   \end{figure}
  
  \subsection{Observations}
  
  At the time of writing, no definitive measurements of the Draconids 2018 have been published. However, as a preliminary comparison with our predictions, the ZHR profiles issued from the International Meteor Organization Visual Meteor Database (IMO/VMDB) and the IMO Video Meteor Network (IMO/VMN)\footnote{\url{https://www.imo.net/draconids-outburst-on-oct-8-9/} accessed on 15/12/2018} were superimposed on the simulation results in Figure \ref{fig:2018}. The black solid curves represent the original measurements, and the light boxes the data re-binned in time steps consistent with the simulated profile (dark boxes). We observe that the length of the simulated profile matches well both sets of observations and that the peak time estimate is accurate to within a half an hour of uncertainty (depending on when the maximum activity really occurred). The simulated intensity underestimates the real strength of the shower, but remains correct within a factor of 2. From these preliminary observations, we conclude that despite the low number of particles involved in the shower simulation, our model led to one of the best predictions of the 2018 Draconids published \citep{Egal2018}. 
  
 \section{Future Draconid outbursts}  \label{sec:future_outbursts}
 
 Finally, in an effort to predict Draconid activity over the next decade, we integrated a sample of particles generated by Model I until the year 2030. In total, \zcorr{360 000 } particles equally spread over our three size bins ($[10^{-4},10^{-3}]$ m, $[10^{-3},10^{-2}]$ m and $[10^{-2},10^{-1}]$ m) were ejected at each apparition of the comet from 1850 to 2030, resulting in a total of about \zcorr{9.72 } million simulated meteoroids. The model predicted overall peak intensity of the shower covering the period 2020-2030 is presented in Figure \ref{fig:years_activity}. 
 
 From this annual variation in the predicted shower strength, we estimate enhanced radio activity in 2019 and 2025, as well as a potential minor display in 2029. Comparing to the predictions of other modelers, we observe that enhanced activity in 2019 and 2025 is also predicted by \cite{Ye2014} and \cite{Maslov2011}. We however do not share the forecast of a low intensity shower in 2021 mentioned in \cite{Ye2014}.

 \subsection{2019}
 
  On October 8 2019, the model predicted that the Earth will face a similar situation to that in 1999 and encounter the streams ejected by 21P in 1959 and 1966. Our estimated activity profile, presented in Figure \ref{fig:ZHR_profiles}, is exclusively composed of particles in the size bin $[10^{-4},10^{-3}]$ m. As with the radio outburst in 1999, a small number of unphysically heavy weighted particles belonging to the 1966 trail raised the ZHR maximum to a storm value of 5000 meteors per hour and have been removed from the analysis.
  
  The low temporal resolution of the simulated profile, typical of our other simulated radio showers, decreases the precision of the time and intensity predictions compared to strong visual outbursts. We  estimate, however, that the Earth might experience a radio outburst with a maximum ZHR of about 200 (within a factor of 3), and enhanced activity between 5h45 and 15h30 UT on October 8. The peak time is difficult to assess from the model profile. If the maximum ZHR is reached around 14h35 UT, ($L_\odot=194.753\degree$), the weighted average of the profile has a peak time of 12h UT ($L_\odot=194.648\degree$). When considering the date of the closest approach between the Earth and the median location of a specific stream, we estimate a maximum activity caused by the 1959 stream around 13h42 UT ($L_\odot=194.718\degree$) and a maximum caused by the 1966 stream at 12h01 UT. The total duration of the shower, among the longest for radio showers in our simulations, is 9h45. 
  
  Simulations performed by other authors differ from our predictions, either in time and intensity. Results from the MSFC model, differing from Model I mostly on the ejection circumstances and the weighting process, augurs for noticeable Draconid activity (nearly as strong as in 2018) culminating around 15h06 UT on October 8 ($L_\odot=194.775\degree$) consisting of mainly smaller particles. A secondary peak might be expected much earlier, around 4h20 the same day ($L_\odot=194.332\degree$). According to \cite{Maslov2011}, the activity level from the 1959 trail in 2019 should be about half of that in 1999. Based on the 1999 ZHR estimated by \cite[Iiyama in][]{Sato2003}, Maslov expects ``a small visual peak not higher than ZHR 5-10, on 2019 October 8 at 14h45m UT''. He however points out that ``radio observations could show much higher activity'' \citep{Maslov2011}. 
  
  In \cite{Jenniskens2006}, no enhanced Draconid activity was predicted by J. Vaubaillon in 2019. However, some revised predictions are presented in Table \ref{table:future_events}. The first one points toward a potential storm on October 8, with a peak time consistent with the MSFC and \cite{Maslov2011}'s results. His ZHR estimate of about 4000 meteors per hour is also in agreement with our simulations when we do not remove the heavily weighted particles. The second prediction, favored by that author because it uses the more accurate ephemeris of the comet provided by JPL, indicates a moderate outburst on October 7 around 11h UT (Vaubaillon, personal communication).
  
   \begin{table*}[!ht]
    \resizebox{\textwidth}{!}{
    \begin{tabular}{lcccccl}
    \hline
    \hline
     Modeler & Trail & Date (y/m/d) & hour &  $L_\odot$ & ZHR & Comment \\
    \hline 
    Egal (1) & Mult. & \zcorr{2019/10/08} & 12h  & 194.648  & 200 $\divideontimes$ 3 & From the ZHR profile (main prediction)\\
      & Mult. & \zcorr{2019/10/08} & 14h35  & 194.753 & 200 $\divideontimes$ 3 & Maximum of the ZHR profile\\
      & 1959 & \zcorr{2019/10/08} & 13h42  & 194.718  & - & Based on stream median location\\
      & 1966 & \zcorr{2019/10/08} & 12h01  & 194.648  & - & Based on stream median location\\
      \hline
    MSFC (2) & Mult. & \zcorr{2019/10/08} & 15h06 & 194.775 & $\sim$ to 2018 & Primary peak \\   
     & Mult. & \zcorr{2019/10/08} & 4h20 & 194.332 & - & Secondary peak \\ 
         \hline 
   Vaubaillon (3) & Mult. & \zcorr{2019/10/08} & 14h41 & 193.522 & 4042 & Old comet ephemeris\\  
     & Mult. & \zcorr{2019/10/07} & 11h01 & 192.811 & 363 & Updated comet ephemeris (recommended)\\
         \hline               
    Maslov (4) & 1959 & \zcorr{2019/10/08} & 14h44 & 194.759 & 5-10 (visual) & Likely higher activity in radio range\\
       \hline  
     Ye (5) & 1979 & & & 194.2 ($\in[194.0,194.5]$) & & Lower rate than in 2018 \\
    \hline
    \end{tabular}}
    \caption{\label{table:future_events} 2019 Draconid forecasts performed by independent modelers. Sources: (1) This work, (2) \cite{Moser2018}, (3) Personal comment, (4) \cite{Maslov2011}, (5) \cite{Ye2014}}
    \end{table*}     
  
  In \cite{Kastinen2017}, no specific activity is discernible for 2019 compared to historic Draconid outbursts. \cite{Ye2014} attribute  enhanced activity in 2019 to small particles belonging to the 1979 trail. The maximum activity is expected to be lower than in 2018 and to peak around $L_\odot=194.2\degree$ ($\in [194,195]\degree$).
 
 A summary of the 2019 predictions is provided in Table \ref{table:future_events}. Different estimates of the shower's characteristics are highly variable and no general agreement about the peak time and intensity is found. When considering the weighted average of our simulated activity profile, the peak time is driven by the influence of the 1966 trail and the resulting prediction is in advance of three hours compared to the MSFC or the \cite{Maslov2011} results. The maximum of the ZHR distribution is consistent with these two models, but was not used in our model's validation. The expectations of the shower's strength are even more uncertain, and vary from a low activity \citep{Maslov2011,Ye2014,Kastinen2017} to a radio outburst (this work, Model I and MSFC model).

 \subsection{2025}
 
 In our simulations, the Earth crosses a portion of the meteoroid stream composed of small particles at the 2025 Draconid return mainly ejected during the 2011-2012 apparition, with some contribution from the 2005 and potentially the 1999 trails. From the activity profile, we estimate a maximum activity around 16h18 UT on October 8 ($L_\odot=195.286$). The Earth is expected to approach the median location of the 2012 trail around 15h30 UT ($L_\odot=195.252\degree$) and the 2005 trail at 13h40 UT ($L_\odot=195.178\degree$). Our maximum intensity ($\divideontimes$ 3) reaches a ZHR of 950 meteors per hour. 
 
 According to \cite{Maslov2011}, encounters with the 1907 to 1953 trails will lead to multiple submaxima at 05h01 UT (ZHR: 10-15), 07h25 UT (ZHR: 20-25), 09h06 UT (ZHR: 20-25), and 10h17 to 10h49 UT (ZHR: 50-60), with many fireballs expected. Consistent with our prediction, \cite{Maslov2011} also predicts an encounter with the 2012 trail on October 8 2025 around 15h14 UT. If the expected visual activity is low (10-40 in ZHR), the author anticipates a much higher radio activity going up to tens of thousand of meteors per hour. 
 Because both forecasts point toward a very strong radio outburst/storm caused by an encounter with the young (and therefore probably dense) 2012 trail, special attention should be paid to this event.  
 
 \subsection{2029}
 
 The 2029 ZHR profile and node plot (cf. Figures \ref{fig:ZHR_profiles} and \ref{fig:nodes}) are hard to interpret. Only some particles ($Npart < 10$) ejected in 1940 approach the Earth in our simulations, leading to the low-resolution shape of Figure \ref{fig:ZHR_profiles}. For this specific shower, more simulations of the 1940 trail in particular are needed. Our first impressions are that we might detect some weak radio activity the night between October 7 and 8, which was not predicted by \cite{Ye2014} or \cite{Maslov2011}.  
 
 \section{Discussion} \label{sec:discussion}
   
  Numerical models of meteor showers are forced to rely on multiple assumptions. Each attempt to better reproduce the comet's behavior requires the incorporation of new parameters, which are usually not measured for the parent body in question. In this section, we address the influence that certain of our assumptions have on our forecasts of future Draconid outbursts. 
 
 \subsection{Comet characteristics and ephemeris}
 
 As mentioned in Section \ref{subsec:21P}, after one century of observations, comet 21P/Giacobini-Zinner remains a mysterious comet. Its physical characteristics are not well defined, and the accuracy of its ephemeris is poor prior to 1966 because of multiple close encounters with Jupiter and erratic variations in its nongravitational forces. A significant orbital discontinuity was observed in 1959-1965, changing the comet ejection patterns and orbital evolution. These considerations led us to adopt orbital solutions available for 21P based on observations at a particular apparition instead of computing its ephemeris backward in time from a precise and recent solution. This decision, forcing us to sometimes rely on less precise orbital measurements (but allowing us to take the 1959-1965 discontinuity into account), may explain the preponderance of different streams involved in our simulated 2005 and 2019 showers compared to other authors \citep{Campbell-Brown2006,Ye2014}. As noticed in \cite{Vaubaillon2011}, the choice of the ephemeris solution has, therefore, an impact on the shower predictions.
 
 By comparing the post-predictions performed using Model I and the MSFC model, which is similar in methodology but assumes different physical characteristics of the comet and ejection circumstances, we observe that both approaches provided consistent conclusions about the years of enhanced Draconid activity, the streams contributing to the shower, and the particles' characteristics (size and ejection velocity). We also notice that the peak time estimates of the historic Draconid outbursts predicted by Model I (from the profiles and the stream median location) and by the MSFC model are similar. We therefore consider the exact choice of the comet's physical parameters and the ejection circumstances to have only a moderate influence in the predictions of the existence and date of enhanced Draconid activity. 

 \subsection{Weights and ZHR computation}

  The weighting solution for comet 21P, detailed in \ref{sec:comet_photometry} and \ref{sec:weights_eq}, has an extremely significant impact on the shower's predicted level of activity. In this work, we have updated the weighting solution of \cite{Vaubaillon2005} considering pre- and post-perihelion dust measurements of comet 21P during the 2018 apparition. The heliocentric distance dependence of the amount of dust ejected by the comet was established from these observations and applied to all the previous apparitions considered in our model.
  
  Despite the evident limitations of this approach, this solution offers a first approximation of the comet's behavior throughout its period of activity and has the merit of being based on real measurements of the comet's activity. Any further modification of 21P's $Af\rho=f(r_h)$ can also easily be implemented in the model to improve future shower predictions. 
  
  Among the unknown parameters of the weighting function \ref{eq:weights}, the most influential and uncertain one is the particle size distribution index at ejection, $u$. Coupled with the number of particles required to produce a ZHR profile ($>15$) and the maximum sphere size adopted for the flux computation ($6h\times V_\oplus$), the value of $u$ can substantially modify the shape, peak time, and intensity of the simulated profiles. It is therefore difficult to conclude that the profiles presented in Figure \ref{fig:ZHR_profiles} provide an accurate prediction of future Draconid outbursts. However, by calibrating these parameters to strong and well-known visual outbursts (cf. Section \ref{sec:model1_calib}), their impact on the final prediction is reduced. 
  
  To illustrate this, we compare the conclusions of the present work with the previous analysis published in \cite{Egal2018}. We observe that with a different weighting solution, a different size distribution index, a higher number of particles required for the profile computation (15 instead of 10), and a larger sample of particles simulated, the accuracy of the time and intensity predictions over the period 1933-2018 are still comparable within uncertainties between the two works. The incorporation of the comet's $Af\rho$ evolution into the model allowed \cite{Egal2018} to obtain correct time and ZHR estimates for historic Draconid outbursts and to perform a high-fidelity prediction of the 2018 outburst (cf. Section \ref{sec:2018}). In this work, the incorporation within the weighting solution of new measurements of the comet's activity (including post-perihelion observations) allowed an increase in the accuracy of our modeling. At the same time the calibration process maintained  consistency with our previous work despite significant variations of our ZHR profile computation. 
 
  \subsubsection{Potential 2019 outburst}
  
  As described in Section \ref{sec:future_outbursts}, a prediction for the 2019 Draconids was particularly difficult to establish and our results diverge from other studies. The low resolution of the simulated activity profile hampers a reliable peak time estimate, while the maximum intensity prediction is complicated by a few unphysically heavily-weighted 1966 particles that we chose to remove. Nevertheless, we predict a radio outburst on October 8 around 12h UT; most other authors expect a lower maximum activity later the same day. Radar observations of the 2019 Draconids will therefore be a good opportunity to both test and improve our model. Precise measurements would help to determine the veracity of our choice to limit the contribution of the 1966 stream, as well as the ephemeris computation implemented in our model. 
  
 The 2019 Draconid return will also offer the opportunity to study two of the most problematic trails in the Draconid complex. The 1959 and 1966 streams, both of which cross the Earth's trajectory in 2019, previously encountered the planet in 1999 and caused a strong radio outburst that was unnoticed until this work. They also intercepted the Earth in 2012, producing a radio storm that was unpredicted by every shower modeler. The fact that 21P suffered an orbital discontinuity precisely during the period 1959-1965 might also point to the activation of local source regions over the comet's surface \citep{Sekanina1993} and a potential for much higher activity than that considered in this model. Special attention should therefore be paid to observations of the coming shower on October 8 2019.

\section{Conclusion}

In this work, we performed numerical simulations of comet 21P/Giacobini-Zinner and the Draconid meteoroid stream complex following the methodology of \cite{Vaubaillon2005}. We generated a total of \zcorr{about 22 million dust particles, ejected by the comet since 1850. } Particles were integrated until the year 2030, and candidates approaching the Earth were carefully selected and analyzed in order to identify past and future Draconid outbursts. 

To our knowledge, this study is the first to produce ZHR profiles for each simulated Draconid shower and quantitatively estimate the peak time and intensity of all the reported historic outbursts. Our model provides a timing accuracy of order half an hour for the peak times and a ZHR estimate correct to within a factor of 2 (visual showers) or 3 (radio outbursts). These successful post-predictions are the result of an updated weighting scheme of the simulated meteoroids, allowing us to better assess the contribution of individual particles to the shower. The initial algorithm of \cite{Vaubaillon2005} has been updated and modified to take into account the evolution of the parent comet dust production along its orbit, adjusted using recent measurements of 21P performed by the NASA Meteoroid Environment Office between May and December 2018. Thanks to this updated solution, we were able to uncover the existence of a previously unreported strong radio outburst in 1999 that we confirmed by archival radar measurements at the predicted date and time. A first set of simulations issued from this model also led to one of the best published predictions of the recent 2018 outburst \citep{Egal2018}. 

In the next decade, we can expect up to three Draconid outbursts caused by radar-sized meteoroids, respectively in 2019, 2025 and potentially 2029. Observations of the 2019 outburst, which we find is among the most challenging of all returns to characterize from our simulations, should particularly help in improving our stream modeling for the Draconid returns of the 2020s. 

From our model predictions we expect a strong radio outburst on October 8 2025 and a somewhat weaker shower in 2019. We encourage observers to be particularly vigilant in recording the Draconid shower in these years.

\section*{Acknowledgements}

We are thankful to J. Vaubaillon for his support and advice regarding his model's implementation. This work was supported in part by NASA Meteoroid Environment Office under cooperative agreement 80NSSC18M0046 and contract 80MSFC18C0011.

\section*{References}

\bibliographystyle{model2-names.bst}\biboptions{authoryear}
\nocite{*}
\bibliography{Draconids}

\begin{thebibliography}{101}
\expandafter\ifx\csname natexlab\endcsname\relax\def\natexlab#1{#1}\fi
\providecommand{\url}[1]{\texttt{#1}}
\providecommand{\href}[2]{#2}
\providecommand{\path}[1]{#1}
\providecommand{\DOIprefix}{doi:}
\providecommand{\ArXivprefix}{arXiv:}
\providecommand{\URLprefix}{URL: }
\providecommand{\Pubmedprefix}{pmid:}
\providecommand{\doi}[1]{\href{http://dx.doi.org/#1}{\path{#1}}}
\providecommand{\Pubmed}[1]{\href{pmid:#1}{\path{#1}}}
\providecommand{\bibinfo}[2]{#2}
\ifx\xfnm\relax \def\xfnm[#1]{\unskip,\space#1}\fi
\bibitem[{{Abramowitz} and {Stegun}(1972)}]{Abramowitz1972}
\bibinfo{author}{{Abramowitz}, M.}, \bibinfo{author}{{Stegun}, I.A.},
  \bibinfo{year}{1972}.
\newblock \bibinfo{title}{{Handbook of Mathematical Functions}}.
\bibitem[{A'Hearn et~al.(1984)A'Hearn, Schleicher, Feldman, Millis and
  Thompson}]{AHearn1984}
\bibinfo{author}{A'Hearn, M.F.}, \bibinfo{author}{Schleicher, D.G.},
  \bibinfo{author}{Feldman, P.D.}, \bibinfo{author}{Millis, R.L.},
  \bibinfo{author}{Thompson, D.T.}, \bibinfo{year}{1984}.
\newblock \bibinfo{title}{{Comet Bowell 1980b}}.
\newblock \bibinfo{journal}{The Astronomical Journal} \bibinfo{volume}{89}.
\bibitem[{{Arlt}(1998)}]{Arlt1998}
\bibinfo{author}{{Arlt}, R.}, \bibinfo{year}{1998}.
\newblock \bibinfo{title}{{Summary of 1998 Draconid Outburst Observations}}.
\newblock \bibinfo{journal}{WGN, Journal of the International Meteor
  Organization} \bibinfo{volume}{26}, \bibinfo{pages}{256--259}.
\bibitem[{{Arlt} et~al.(1999){Arlt}, {Bellot Rubio}, {Brown} and
  {Gyssens}}]{Arlt1999}
\bibinfo{author}{{Arlt}, R.}, \bibinfo{author}{{Bellot Rubio}, L.},
  \bibinfo{author}{{Brown}, P.}, \bibinfo{author}{{Gyssens}, M.},
  \bibinfo{year}{1999}.
\newblock \bibinfo{title}{{Bulletin 15 of the International Leonid Watch: First
  Global Analysis of the 1999 Leonid Storm}}.
\newblock \bibinfo{journal}{WGN, Journal of the International Meteor
  Organization} \bibinfo{volume}{27}, \bibinfo{pages}{286--295}.
\bibitem[{{Astropy Collaboration} et~al.(2013){Astropy Collaboration},
  {Robitaille}, {Tollerud}, {Greenfield}, {Droettboom}, {Bray}, {Aldcroft},
  {Davis}, {Ginsburg}, {Price-Whelan}, {Kerzendorf}, {Conley}, {Crighton},
  {Barbary}, {Muna}, {Ferguson}, {Grollier}, {Parikh}, {Nair}, {Unther},
  {Deil}, {Woillez}, {Conseil}, {Kramer}, {Turner}, {Singer}, {Fox}, {Weaver},
  {Zabalza}, {Edwards}, {Azalee Bostroem}, {Burke}, {Casey}, {Crawford},
  {Dencheva}, {Ely}, {Jenness}, {Labrie}, {Lim}, {Pierfederici}, {Pontzen},
  {Ptak}, {Refsdal}, {Servillat} and {Streicher}}]{Astropy2013}
\bibinfo{author}{{Astropy Collaboration}}, \bibinfo{author}{{Robitaille},
  T.P.}, \bibinfo{author}{{Tollerud}, E.J.}, \bibinfo{author}{{Greenfield},
  P.}, \bibinfo{author}{{Droettboom}, M.}, \bibinfo{author}{{Bray}, E.},
  \bibinfo{author}{{Aldcroft}, T.}, \bibinfo{author}{{Davis}, M.},
  \bibinfo{author}{{Ginsburg}, A.}, \bibinfo{author}{{Price-Whelan}, A.M.},
  \bibinfo{author}{{Kerzendorf}, W.E.}, \bibinfo{author}{{Conley}, A.},
  \bibinfo{author}{{Crighton}, N.}, \bibinfo{author}{{Barbary}, K.},
  \bibinfo{author}{{Muna}, D.}, \bibinfo{author}{{Ferguson}, H.},
  \bibinfo{author}{{Grollier}, F.}, \bibinfo{author}{{Parikh}, M.M.},
  \bibinfo{author}{{Nair}, P.H.}, \bibinfo{author}{{Unther}, H.M.},
  \bibinfo{author}{{Deil}, C.}, \bibinfo{author}{{Woillez}, J.},
  \bibinfo{author}{{Conseil}, S.}, \bibinfo{author}{{Kramer}, R.},
  \bibinfo{author}{{Turner}, J.E.H.}, \bibinfo{author}{{Singer}, L.},
  \bibinfo{author}{{Fox}, R.}, \bibinfo{author}{{Weaver}, B.A.},
  \bibinfo{author}{{Zabalza}, V.}, \bibinfo{author}{{Edwards}, Z.I.},
  \bibinfo{author}{{Azalee Bostroem}, K.}, \bibinfo{author}{{Burke}, D.J.},
  \bibinfo{author}{{Casey}, A.R.}, \bibinfo{author}{{Crawford}, S.M.},
  \bibinfo{author}{{Dencheva}, N.}, \bibinfo{author}{{Ely}, J.},
  \bibinfo{author}{{Jenness}, T.}, \bibinfo{author}{{Labrie}, K.},
  \bibinfo{author}{{Lim}, P.L.}, \bibinfo{author}{{Pierfederici}, F.},
  \bibinfo{author}{{Pontzen}, A.}, \bibinfo{author}{{Ptak}, A.},
  \bibinfo{author}{{Refsdal}, B.}, \bibinfo{author}{{Servillat}, M.},
  \bibinfo{author}{{Streicher}, O.}, \bibinfo{year}{2013}.
\newblock \bibinfo{title}{{Astropy: A community Python package for astronomy}}.
\newblock \bibinfo{journal}{Astronomy \& Astrophysics} \bibinfo{volume}{558},
  \bibinfo{pages}{A33}.
\newblock \DOIprefix\doi{10.1051/0004-6361/201322068},
  \href{http://arxiv.org/abs/1307.6212}{\tt arXiv:1307.6212}.
\bibitem[{{Blaauw} et~al.(2011){Blaauw}, {Campbell-Brown} and
  {Weryk}}]{Blaauw2011}
\bibinfo{author}{{Blaauw}, R.C.}, \bibinfo{author}{{Campbell-Brown}, M.D.},
  \bibinfo{author}{{Weryk}, R.J.}, \bibinfo{year}{2011}.
\newblock \bibinfo{title}{{A meteoroid stream survey using the Canadian Meteor
  Orbit Radar - III. Mass distribution indices of six major meteor showers}}.
\newblock \bibinfo{journal}{Monthly Notices of the Royal Astronomical Society}
  \bibinfo{volume}{414}, \bibinfo{pages}{3322--3329}.
\newblock \DOIprefix\doi{10.1111/j.1365-2966.2011.18633.x}.
\bibitem[{Blaauw et~al.(2014)Blaauw, Suggs and Cooke}]{Blaauw2014}
\bibinfo{author}{Blaauw, R.C.}, \bibinfo{author}{Suggs, R.M.},
  \bibinfo{author}{Cooke, W.J.}, \bibinfo{year}{2014}.
\newblock \bibinfo{title}{{Dust Production of Comet 21P/Giacobini-Zinner using
  Broadband Photometry}}.
\newblock \bibinfo{journal}{Meteoritics {\&} Planetary Science}
  \bibinfo{volume}{49}, \bibinfo{pages}{1--10}.
\bibitem[{{Borovi{\v c}ka} et~al.(2007){Borovi{\v c}ka}, {Spurn{\'y}} and
  {Koten}}]{Borovicka2007}
\bibinfo{author}{{Borovi{\v c}ka}, J.}, \bibinfo{author}{{Spurn{\'y}}, P.},
  \bibinfo{author}{{Koten}, P.}, \bibinfo{year}{2007}.
\newblock \bibinfo{title}{{Atmospheric deceleration and light curves of
  Draconid meteors and implications for the structure of cometary dust}}.
\newblock \bibinfo{journal}{Astronomy and Astrophysics} \bibinfo{volume}{473},
  \bibinfo{pages}{661--672}.
\newblock \DOIprefix\doi{10.1051/0004-6361:20078131}.
\bibitem[{{Brown} et~al.(2000){Brown}, {Campbell}, {Ellis}, {Hawkes}, {Jones},
  {Gural}, {Babcock}, {Barnbaum}, {Bartlett}, {Bedard}, {Bedient}, {Beech},
  {Brosch}, {Clifton}, {Connors}, {Cooke}, {Goetz}, {Gaines}, {Gramer}, {Gray},
  {Hildebrand}, {Jewell}, {Jones}, {Leake}, {LeBlanc}, {Looper}, {McIntosh},
  {Montague}, {Morrow}, {Murray}, {Nikolova}, {Robichaud}, {Spondor},
  {Talarico}, {Theijsmeijer}, {Tilton}, {Treu}, {Vachon}, {Webster}, {Weryk}
  and {Worden}}]{Brown2000}
\bibinfo{author}{{Brown}, P.}, \bibinfo{author}{{Campbell}, M.D.},
  \bibinfo{author}{{Ellis}, K.J.}, \bibinfo{author}{{Hawkes}, R.L.},
  \bibinfo{author}{{Jones}, J.}, \bibinfo{author}{{Gural}, P.},
  \bibinfo{author}{{Babcock}, D.}, \bibinfo{author}{{Barnbaum}, C.},
  \bibinfo{author}{{Bartlett}, R.K.}, \bibinfo{author}{{Bedard}, M.},
  \bibinfo{author}{{Bedient}, J.}, \bibinfo{author}{{Beech}, M.},
  \bibinfo{author}{{Brosch}, N.}, \bibinfo{author}{{Clifton}, S.},
  \bibinfo{author}{{Connors}, M.}, \bibinfo{author}{{Cooke}, B.},
  \bibinfo{author}{{Goetz}, P.}, \bibinfo{author}{{Gaines}, J.K.},
  \bibinfo{author}{{Gramer}, L.}, \bibinfo{author}{{Gray}, J.},
  \bibinfo{author}{{Hildebrand}, A.R.}, \bibinfo{author}{{Jewell}, D.},
  \bibinfo{author}{{Jones}, A.}, \bibinfo{author}{{Leake}, M.},
  \bibinfo{author}{{LeBlanc}, A.G.}, \bibinfo{author}{{Looper}, J.K.},
  \bibinfo{author}{{McIntosh}, B.A.}, \bibinfo{author}{{Montague}, T.},
  \bibinfo{author}{{Morrow}, M.J.}, \bibinfo{author}{{Murray}, I.S.},
  \bibinfo{author}{{Nikolova}, S.}, \bibinfo{author}{{Robichaud}, J.},
  \bibinfo{author}{{Spondor}, R.}, \bibinfo{author}{{Talarico}, J.},
  \bibinfo{author}{{Theijsmeijer}, C.}, \bibinfo{author}{{Tilton}, B.},
  \bibinfo{author}{{Treu}, M.}, \bibinfo{author}{{Vachon}, C.},
  \bibinfo{author}{{Webster}, A.R.}, \bibinfo{author}{{Weryk}, R.},
  \bibinfo{author}{{Worden}, S.P.}, \bibinfo{year}{2000}.
\newblock \bibinfo{title}{{Global Ground-Based Electro-Optical and Radar
  Observations of the 1999 Leonid Shower: First Results}}.
\newblock \bibinfo{journal}{Earth Moon and Planets} \bibinfo{volume}{82},
  \bibinfo{pages}{167--190}.
\bibitem[{{Brown} and {Jones}(1998)}]{Brown1998}
\bibinfo{author}{{Brown}, P.}, \bibinfo{author}{{Jones}, J.},
  \bibinfo{year}{1998}.
\newblock \bibinfo{title}{{Simulation of the Formation and Evolution of the
  Perseid Meteoroid Stream}}.
\newblock \bibinfo{journal}{Icarus} \bibinfo{volume}{133},
  \bibinfo{pages}{36--68}.
\newblock \DOIprefix\doi{10.1006/icar.1998.5920}.
\bibitem[{{Campbell-Brown} et~al.(2006){Campbell-Brown}, {Vaubaillon}, {Brown},
  {Weryk} and {Arlt}}]{Campbell-Brown2006}
\bibinfo{author}{{Campbell-Brown}, M.}, \bibinfo{author}{{Vaubaillon}, J.},
  \bibinfo{author}{{Brown}, P.}, \bibinfo{author}{{Weryk}, R.J.},
  \bibinfo{author}{{Arlt}, R.}, \bibinfo{year}{2006}.
\newblock \bibinfo{title}{{The 2005 Draconid outburst}}.
\newblock \bibinfo{journal}{Astronomy and Astrophysics} \bibinfo{volume}{451},
  \bibinfo{pages}{339--344}.
\newblock \DOIprefix\doi{10.1051/0004-6361:20054588}.
\bibitem[{{Chebotarev} and {Simek}(1987)}]{Chebotarev1987}
\bibinfo{author}{{Chebotarev}, R.P.}, \bibinfo{author}{{Simek}, M.},
  \bibinfo{year}{1987}.
\newblock \bibinfo{title}{{The structure of the Giacobinid 1985 meteor shower
  from radar observations in Dushanbe and Ondrejov}}.
\newblock \bibinfo{journal}{Bulletin of the Astronomical Institutes of
  Czechoslovakia} \bibinfo{volume}{38}, \bibinfo{pages}{362--367}.
\bibitem[{Churyumov and Rosenbush(1991)}]{Churyumov1991}
\bibinfo{author}{Churyumov, K.I.}, \bibinfo{author}{Rosenbush, V.K.},
  \bibinfo{year}{1991}.
\newblock \bibinfo{title}{{Peculiarities of gas and dust production rates in
  comets P/Halley (1986III), P/Giacobini-Zinner (1985 XIII), P/Hartley-Good
  (1985 XVII) and P/Thiele (1985XIX)}}.
\newblock \bibinfo{journal}{Astron. Nachr} \bibinfo{volume}{312},
  \bibinfo{pages}{385--391}.
\bibitem[{{Cook}(1973)}]{Cook1973}
\bibinfo{author}{{Cook}, A.F.}, \bibinfo{year}{1973}.
\newblock \bibinfo{title}{{A Working List of Meteor Streams}}.
\newblock \bibinfo{journal}{NASA Special Publication} \bibinfo{volume}{319},
  \bibinfo{pages}{183}.
\bibitem[{{Crifo} and {Rodionov}(1997)}]{Crifo1997}
\bibinfo{author}{{Crifo}, J.F.}, \bibinfo{author}{{Rodionov}, A.V.},
  \bibinfo{year}{1997}.
\newblock \bibinfo{title}{{The Dependence of the Circumnuclear Coma Structure
  on the Properties of the Nucleus}}.
\newblock \bibinfo{journal}{Icarus} \bibinfo{volume}{127},
  \bibinfo{pages}{319--353}.
\newblock \DOIprefix\doi{10.1006/icar.1997.5690}.
\bibitem[{{Davies} and {Lovell}(1955)}]{Davies1955}
\bibinfo{author}{{Davies}, J.G.}, \bibinfo{author}{{Lovell}, A.C.B.},
  \bibinfo{year}{1955}.
\newblock \bibinfo{title}{{The Giacobinid meteor stream}}.
\newblock \bibinfo{journal}{Monthly Notices of the RAS} \bibinfo{volume}{115},
  \bibinfo{pages}{23}.
\newblock \DOIprefix\doi{10.1093/mnras/115.1.23}.
\bibitem[{{Della Corte, V.} et~al.(2015){Della Corte, V.}, {Rotundi, A.},
  {Fulle, M.}, {Gruen, E.}, {Weissman, P.}, {Sordini, R.}, {Ferrari, M.},
  {Ivanovski, S.}, {Lucarelli, F.}, {Accolla, M.}, {Zakharov, V.}, {Mazzotta
  Epifani, E.}, {Lopez-Moreno, J. J.}, {Rodriguez, J.}, {Colangeli, L.},
  {Palumbo, P.}, {Bussoletti, E.}, {Crifo, J. F.}, {Esposito, F.}, {Green, S.
  F.}, {Lamy, P. L.}, {McDonnell, J. A. M.}, {Mennella, V.}, {Molina, A.},
  {Morales, R.}, {Moreno, F.}, {Ortiz, J. L.}, {Palomba, E.}, {Perrin, J. M.},
  {Rietmeijer, F. J. M.}, {Rodrigo, R.}, {Zarnecki, J. C.}, {Cosi, M.},
  {Giovane, F.}, {Gustafson, B.}, {Herranz, M. L.}, {Jeronimo, J. M.}, {Leese,
  M. R.}, {Lopez-Jimenez, A. C.} and {Altobelli, N.}}]{DellaCorte2015}
\bibinfo{author}{{Della Corte, V.}}, \bibinfo{author}{{Rotundi, A.}},
  \bibinfo{author}{{Fulle, M.}}, \bibinfo{author}{{Gruen, E.}},
  \bibinfo{author}{{Weissman, P.}}, \bibinfo{author}{{Sordini, R.}},
  \bibinfo{author}{{Ferrari, M.}}, \bibinfo{author}{{Ivanovski, S.}},
  \bibinfo{author}{{Lucarelli, F.}}, \bibinfo{author}{{Accolla, M.}},
  \bibinfo{author}{{Zakharov, V.}}, \bibinfo{author}{{Mazzotta Epifani, E.}},
  \bibinfo{author}{{Lopez-Moreno, J. J.}}, \bibinfo{author}{{Rodriguez, J.}},
  \bibinfo{author}{{Colangeli, L.}}, \bibinfo{author}{{Palumbo, P.}},
  \bibinfo{author}{{Bussoletti, E.}}, \bibinfo{author}{{Crifo, J. F.}},
  \bibinfo{author}{{Esposito, F.}}, \bibinfo{author}{{Green, S. F.}},
  \bibinfo{author}{{Lamy, P. L.}}, \bibinfo{author}{{McDonnell, J. A. M.}},
  \bibinfo{author}{{Mennella, V.}}, \bibinfo{author}{{Molina, A.}},
  \bibinfo{author}{{Morales, R.}}, \bibinfo{author}{{Moreno, F.}},
  \bibinfo{author}{{Ortiz, J. L.}}, \bibinfo{author}{{Palomba, E.}},
  \bibinfo{author}{{Perrin, J. M.}}, \bibinfo{author}{{Rietmeijer, F. J. M.}},
  \bibinfo{author}{{Rodrigo, R.}}, \bibinfo{author}{{Zarnecki, J. C.}},
  \bibinfo{author}{{Cosi, M.}}, \bibinfo{author}{{Giovane, F.}},
  \bibinfo{author}{{Gustafson, B.}}, \bibinfo{author}{{Herranz, M. L.}},
  \bibinfo{author}{{Jeronimo, J. M.}}, \bibinfo{author}{{Leese, M. R.}},
  \bibinfo{author}{{Lopez-Jimenez, A. C.}}, \bibinfo{author}{{Altobelli, N.}},
  \bibinfo{year}{2015}.
\newblock \bibinfo{title}{Giada: shining a light on the monitoring of the comet
  dust production from the nucleus of 67p/churyumov-gerasimenko}.
\newblock \bibinfo{journal}{A\&A} \bibinfo{volume}{583}, \bibinfo{pages}{A13}.
\newblock \URLprefix \url{https://doi.org/10.1051/0004-6361/201526208},
  \DOIprefix\doi{10.1051/0004-6361/201526208}.
\bibitem[{{Denning}(1927)}]{Denning1927}
\bibinfo{author}{{Denning}, W.F.}, \bibinfo{year}{1927}.
\newblock \bibinfo{title}{{La Nouvelle Chute Meteorique Cometaire.}}
\newblock \bibinfo{journal}{L'Astronomie} \bibinfo{volume}{41},
  \bibinfo{pages}{374--376}.
\bibitem[{{Egal} et~al.(2018){Egal}, {Wiegert}, {Brown}, {Moser}, {Moorhead}
  and {Cooke}}]{Egal2018}
\bibinfo{author}{{Egal}, A.}, \bibinfo{author}{{Wiegert}, P.},
  \bibinfo{author}{{Brown}, P.G.}, \bibinfo{author}{{Moser}, D.E.},
  \bibinfo{author}{{Moorhead}, A.V.}, \bibinfo{author}{{Cooke}, W.J.},
  \bibinfo{year}{2018}.
\newblock \bibinfo{title}{{The Draconid Meteoroid Stream 2018: Prospects for
  Satellite Impact Detection}}.
\newblock \bibinfo{journal}{\apjl} \bibinfo{volume}{866}, \bibinfo{pages}{L8}.
\newblock \DOIprefix\doi{10.3847/2041-8213/aae2ba},
  \href{http://arxiv.org/abs/1809.07393}{\tt arXiv:1809.07393}.
\bibitem[{Everhart(1985)}]{Everhart1985}
\bibinfo{author}{Everhart, E.}, \bibinfo{year}{1985}.
\newblock \bibinfo{title}{{An efficient integrator that uses Gauss-Radau
  spacings}}.
\newblock \bibinfo{journal}{International Astronomical Union Colloquium}
  \bibinfo{volume}{83}, \bibinfo{pages}{185–202}.
\bibitem[{{Fisher}(1934)}]{Fisher1934}
\bibinfo{author}{{Fisher}, W.J.}, \bibinfo{year}{1934}.
\newblock \bibinfo{title}{{The History of the Giacobinid Meteors}}.
\newblock \bibinfo{journal}{Harvard College Observatory Bulletin}
  \bibinfo{volume}{894}, \bibinfo{pages}{15--16}.
\bibitem[{{Fujiwara} et~al.(2016){Fujiwara}, {Kero}, {Abo}, {Szasz} and
  {Nakamura}}]{Fujiwara2016}
\bibinfo{author}{{Fujiwara}, Y.}, \bibinfo{author}{{Kero}, J.},
  \bibinfo{author}{{Abo}, M.}, \bibinfo{author}{{Szasz}, C.},
  \bibinfo{author}{{Nakamura}, T.}, \bibinfo{year}{2016}.
\newblock \bibinfo{title}{{MU radar head echo observations of the 2012 October
  Draconid outburst}}.
\newblock \bibinfo{journal}{Monthly Notices of the Royal Astronomical Society,}
  \bibinfo{volume}{455}, \bibinfo{pages}{3273--3280}.
\newblock \DOIprefix\doi{10.1093/mnras/stv2492}.
\bibitem[{{Fulle} et~al.(2016){Fulle}, {Marzari}, {Della Corte}, {Fornasier},
  {Sierks}, {Rotundi}, {Barbieri}, {Lamy}, {Rodrigo}, {Koschny}, {Rickman},
  {Keller}, {L{\'o}pez-Moreno}, {Accolla}, {Agarwal}, {A'Hearn}, {Altobelli},
  {Barucci}, {Bertaux}, {Bertini}, {Bodewits}, {Bussoletti}, {Colangeli},
  {Cosi}, {Cremonese}, {Crifo}, {Da Deppo}, {Davidsson}, {Debei}, {De Cecco},
  {Esposito}, {Ferrari}, {Giovane}, {Gustafson}, {Green}, {Groussin},
  {Gr{\"u}n}, {Gutierrez}, {G{\"u}ttler}, {Herranz}, {Hviid}, {Ip},
  {Ivanovski}, {Jer{\'o}nimo}, {Jorda}, {Knollenberg}, {Kramm}, {K{\"u}hrt},
  {K{\"u}ppers}, {Lara}, {Lazzarin}, {Leese}, {L{\'o}pez-Jim{\'e}nez},
  {Lucarelli}, {Mazzotta Epifani}, {McDonnell}, {Mennella}, {Molina},
  {Morales}, {Moreno}, {Mottola}, {Naletto}, {Oklay}, {Ortiz}, {Palomba},
  {Palumbo}, {Perrin}, {Rietmeijer}, {Rodr{\'{\i}}guez}, {Sordini}, {Thomas},
  {Tubiana}, {Vincent}, {Weissman}, {Wenzel}, {Zakharov} and
  {Zarnecki}}]{Fulle2016}
\bibinfo{author}{{Fulle}, M.}, \bibinfo{author}{{Marzari}, F.},
  \bibinfo{author}{{Della Corte}, V.}, \bibinfo{author}{{Fornasier}, S.},
  \bibinfo{author}{{Sierks}, H.}, \bibinfo{author}{{Rotundi}, A.},
  \bibinfo{author}{{Barbieri}, C.}, \bibinfo{author}{{Lamy}, P.L.},
  \bibinfo{author}{{Rodrigo}, R.}, \bibinfo{author}{{Koschny}, D.},
  \bibinfo{author}{{Rickman}, H.}, \bibinfo{author}{{Keller}, H.U.},
  \bibinfo{author}{{L{\'o}pez-Moreno}, J.J.}, \bibinfo{author}{{Accolla}, M.},
  \bibinfo{author}{{Agarwal}, J.}, \bibinfo{author}{{A'Hearn}, M.F.},
  \bibinfo{author}{{Altobelli}, N.}, \bibinfo{author}{{Barucci}, M.A.},
  \bibinfo{author}{{Bertaux}, J.L.}, \bibinfo{author}{{Bertini}, I.},
  \bibinfo{author}{{Bodewits}, D.}, \bibinfo{author}{{Bussoletti}, E.},
  \bibinfo{author}{{Colangeli}, L.}, \bibinfo{author}{{Cosi}, M.},
  \bibinfo{author}{{Cremonese}, G.}, \bibinfo{author}{{Crifo}, J.F.},
  \bibinfo{author}{{Da Deppo}, V.}, \bibinfo{author}{{Davidsson}, B.},
  \bibinfo{author}{{Debei}, S.}, \bibinfo{author}{{De Cecco}, M.},
  \bibinfo{author}{{Esposito}, F.}, \bibinfo{author}{{Ferrari}, M.},
  \bibinfo{author}{{Giovane}, F.}, \bibinfo{author}{{Gustafson}, B.},
  \bibinfo{author}{{Green}, S.F.}, \bibinfo{author}{{Groussin}, O.},
  \bibinfo{author}{{Gr{\"u}n}, E.}, \bibinfo{author}{{Gutierrez}, P.},
  \bibinfo{author}{{G{\"u}ttler}, C.}, \bibinfo{author}{{Herranz}, M.L.},
  \bibinfo{author}{{Hviid}, S.F.}, \bibinfo{author}{{Ip}, W.},
  \bibinfo{author}{{Ivanovski}, S.L.}, \bibinfo{author}{{Jer{\'o}nimo}, J.M.},
  \bibinfo{author}{{Jorda}, L.}, \bibinfo{author}{{Knollenberg}, J.},
  \bibinfo{author}{{Kramm}, R.}, \bibinfo{author}{{K{\"u}hrt}, E.},
  \bibinfo{author}{{K{\"u}ppers}, M.}, \bibinfo{author}{{Lara}, L.},
  \bibinfo{author}{{Lazzarin}, M.}, \bibinfo{author}{{Leese}, M.R.},
  \bibinfo{author}{{L{\'o}pez-Jim{\'e}nez}, A.C.},
  \bibinfo{author}{{Lucarelli}, F.}, \bibinfo{author}{{Mazzotta Epifani}, E.},
  \bibinfo{author}{{McDonnell}, J.A.M.}, \bibinfo{author}{{Mennella}, V.},
  \bibinfo{author}{{Molina}, A.}, \bibinfo{author}{{Morales}, R.},
  \bibinfo{author}{{Moreno}, F.}, \bibinfo{author}{{Mottola}, S.},
  \bibinfo{author}{{Naletto}, G.}, \bibinfo{author}{{Oklay}, N.},
  \bibinfo{author}{{Ortiz}, J.L.}, \bibinfo{author}{{Palomba}, E.},
  \bibinfo{author}{{Palumbo}, P.}, \bibinfo{author}{{Perrin}, J.M.},
  \bibinfo{author}{{Rietmeijer}, F.J.M.}, \bibinfo{author}{{Rodr{\'{\i}}guez},
  J.}, \bibinfo{author}{{Sordini}, R.}, \bibinfo{author}{{Thomas}, N.},
  \bibinfo{author}{{Tubiana}, C.}, \bibinfo{author}{{Vincent}, J.B.},
  \bibinfo{author}{{Weissman}, P.}, \bibinfo{author}{{Wenzel}, K.P.},
  \bibinfo{author}{{Zakharov}, V.}, \bibinfo{author}{{Zarnecki}, J.C.},
  \bibinfo{year}{2016}.
\newblock \bibinfo{title}{{Evolution of the Dust Size Distribution of Comet
  67P/Churyumov-Gerasimenko from 2.2 au to Perihelion}}.
\newblock \bibinfo{journal}{Astrophysical Journal} \bibinfo{volume}{821},
  \bibinfo{pages}{19}.
\newblock \DOIprefix\doi{10.3847/0004-637X/821/1/19}.
\bibitem[{Green et~al.(2004)Green, McDonnell, McBride, Colwell, Tuzzolino,
  Economou, Tsou, Clark and Brownlee}]{Green2004}
\bibinfo{author}{Green, S.F.}, \bibinfo{author}{McDonnell, J.A.M.},
  \bibinfo{author}{McBride, N.}, \bibinfo{author}{Colwell, M.T.S.H.},
  \bibinfo{author}{Tuzzolino, A.J.}, \bibinfo{author}{Economou, T.E.},
  \bibinfo{author}{Tsou, P.}, \bibinfo{author}{Clark, B.C.},
  \bibinfo{author}{Brownlee, D.E.}, \bibinfo{year}{2004}.
\newblock \bibinfo{title}{The dust mass distribution of comet 81p/wild 2}.
\newblock \bibinfo{journal}{Journal of Geophysical Research: Planets}
  \bibinfo{volume}{109}.
\newblock \DOIprefix\doi{10.1029/2004JE002318}.
\bibitem[{{Hanner} et~al.(1992){Hanner}, {Veeder} and {Tokunaga}}]{Hanner1992}
\bibinfo{author}{{Hanner}, M.S.}, \bibinfo{author}{{Veeder}, G.J.},
  \bibinfo{author}{{Tokunaga}, A.T.}, \bibinfo{year}{1992}.
\newblock \bibinfo{title}{{The dust coma of comet P/Giacobini-Zinner in the
  infrared}}.
\newblock \bibinfo{journal}{The Astronomical Journal} \bibinfo{volume}{104},
  \bibinfo{pages}{386--393}.
\bibitem[{{Hosek} et~al.(2013){Hosek}, {Blaauw}, {Cooke} and
  {Suggs}}]{Hosek2013}
\bibinfo{author}{{Hosek}, Matthew~W., J.}, \bibinfo{author}{{Blaauw}, R.C.},
  \bibinfo{author}{{Cooke}, W.J.}, \bibinfo{author}{{Suggs}, R.M.},
  \bibinfo{year}{2013}.
\newblock \bibinfo{title}{{Outburst Dust Production of Comet
  29P/Schwassmann-Wachmann 1}}.
\newblock \bibinfo{journal}{\aj} \bibinfo{volume}{145}, \bibinfo{pages}{122}.
\newblock \DOIprefix\doi{10.1088/0004-6256/145/5/122}.
\bibitem[{{Hughes} and {Thompson}(1973)}]{Hughes1973}
\bibinfo{author}{{Hughes}, D.W.}, \bibinfo{author}{{Thompson}, D.A.},
  \bibinfo{year}{1973}.
\newblock \bibinfo{title}{{The Giacobinid (Draconid) meteor shower, 1972}}.
\newblock \bibinfo{journal}{Monthly Notices of the RAS} \bibinfo{volume}{163},
  \bibinfo{pages}{3P}.
\newblock \DOIprefix\doi{10.1093/mnras/163.1.3P}.
\bibitem[{{Hutcherson}(1946)}]{Hutcherson1946}
\bibinfo{author}{{Hutcherson}, W.R.}, \bibinfo{year}{1946}.
\newblock \bibinfo{title}{{Spectacular shower of meteors observed in Berea,
  Kentucky}}.
\newblock \bibinfo{journal}{Popular Astronomy} \bibinfo{volume}{54},
  \bibinfo{pages}{484}.
\bibitem[{{Jenniskens}(1995)}]{Jenniskens1995}
\bibinfo{author}{{Jenniskens}, P.}, \bibinfo{year}{1995}.
\newblock \bibinfo{title}{{Meteor stream activity. 2: Meteor outbursts}}.
\newblock \bibinfo{journal}{Astronomy and Astrophysics} \bibinfo{volume}{295},
  \bibinfo{pages}{206--235}.
\bibitem[{{Jenniskens}(2006)}]{Jenniskens2006}
\bibinfo{author}{{Jenniskens}, P.}, \bibinfo{year}{2006}.
\newblock \bibinfo{title}{{Meteor Showers and their Parent Comets}}.
\bibitem[{{Jewitt} and {Meech}(1987)}]{Jewitt1987}
\bibinfo{author}{{Jewitt}, D.C.}, \bibinfo{author}{{Meech}, K.J.},
  \bibinfo{year}{1987}.
\newblock \bibinfo{title}{{Surface brightness profiles of 10 comets}}.
\newblock \bibinfo{journal}{\apj} \bibinfo{volume}{317},
  \bibinfo{pages}{992--1001}.
\newblock \DOIprefix\doi{10.1086/165347}.
\bibitem[{{Jodrell Bank}(1953)}]{Jodrell1953}
\bibinfo{author}{{Jodrell Bank}}, \bibinfo{year}{1953}.
\newblock \bibinfo{title}{{Report on Jodrell Bank to the Royal Astronomical
  Society Council for 1953}}, in: \bibinfo{booktitle}{Proceedings of
  Observatories - Jodrell Bank reprint No 97}.
\bibitem[{{Jones}(1995)}]{Jones1995}
\bibinfo{author}{{Jones}, J.}, \bibinfo{year}{1995}.
\newblock \bibinfo{title}{{The ejection of meteoroids from comets}}.
\newblock \bibinfo{journal}{Monthly Notices of the RAS} \bibinfo{volume}{275},
  \bibinfo{pages}{773--780}.
\newblock \DOIprefix\doi{10.1093/mnras/275.3.773}.
\bibitem[{{Jones} and {Brown}(1996)}]{Jones1996}
\bibinfo{author}{{Jones}, J.}, \bibinfo{author}{{Brown}, P.G.},
  \bibinfo{year}{1996}.
\newblock in: \bibinfo{editor}{{Gustafson}, B.A.S.}, \bibinfo{editor}{{Hanner},
  M.S.} (Eds.), \bibinfo{booktitle}{{Physics, chemistry, and dynamics of
  interplanetary dust}}. volume \bibinfo{volume}{104} of
  \textit{\bibinfo{series}{Astronomical Society of the Pacific Conference
  Series}}.
\bibitem[{{Jorda} et~al.(1992){Jorda}, {Crovisier} and {Green}}]{Jorda1992}
\bibinfo{author}{{Jorda}, L.}, \bibinfo{author}{{Crovisier}, J.},
  \bibinfo{author}{{Green}, D.W.E.}, \bibinfo{year}{1992}.
\newblock \bibinfo{title}{{The correlation between water production rates and
  visual magnitudes in comets}}, in: \bibinfo{editor}{{Harris}, A.W.},
  \bibinfo{editor}{{Bowell}, E.} (Eds.), \bibinfo{booktitle}{Asteroids, Comets,
  Meteors 1991}.
\bibitem[{{Kac}(2015)}]{Kac2015}
\bibinfo{author}{{Kac}, J.}, \bibinfo{year}{2015}.
\newblock \bibinfo{title}{{Draconid 2011 outburst observations from Slovenia}}.
\newblock \bibinfo{journal}{WGN, Journal of the International Meteor
  Organization} \bibinfo{volume}{43}, \bibinfo{pages}{75--80}.
\bibitem[{{Kastinen} and {Kero}(2017)}]{Kastinen2017}
\bibinfo{author}{{Kastinen}, D.}, \bibinfo{author}{{Kero}, J.},
  \bibinfo{year}{2017}.
\newblock \bibinfo{title}{{A Monte Carlo-type simulation toolbox for Solar
  System small body dynamics: Application to the October Draconids}}.
\newblock \bibinfo{journal}{Planetary Space Science} \bibinfo{volume}{143},
  \bibinfo{pages}{53--66}.
\newblock \DOIprefix\doi{10.1016/j.pss.2017.03.007}.
\bibitem[{{Kero} et~al.(2012){Kero}, {Fujiwara}, {Abo}, {Szasz} and
  {Nakamura}}]{Kero2012}
\bibinfo{author}{{Kero}, J.}, \bibinfo{author}{{Fujiwara}, Y.},
  \bibinfo{author}{{Abo}, M.}, \bibinfo{author}{{Szasz}, C.},
  \bibinfo{author}{{Nakamura}, T.}, \bibinfo{year}{2012}.
\newblock \bibinfo{title}{{MU radar head echo observations of the 2011 October
  Draconids}}.
\newblock \bibinfo{journal}{Monthly Notices of the RAS} \bibinfo{volume}{424},
  \bibinfo{pages}{1799--1806}.
\newblock \DOIprefix\doi{10.1111/j.1365-2966.2012.21255.x}.
\bibitem[{{Koschack} and {Rendtel}(1990)}]{Koschack1990}
\bibinfo{author}{{Koschack}, R.}, \bibinfo{author}{{Rendtel}, J.},
  \bibinfo{year}{1990}.
\newblock \bibinfo{title}{{Determination of spatial number density and mass
  index from visual meteor observations (I).}}
\newblock \bibinfo{journal}{WGN, Journal of the International Meteor
  Organization} \bibinfo{volume}{18}, \bibinfo{pages}{44--58}.
\bibitem[{{Koseki}(1990)}]{Koseki1990}
\bibinfo{author}{{Koseki}, M.}, \bibinfo{year}{1990}.
\newblock \bibinfo{title}{{Observations of the 1985 Giacobinids in Japan}}.
\newblock \bibinfo{journal}{Icarus} \bibinfo{volume}{88},
  \bibinfo{pages}{122--128}.
\newblock \DOIprefix\doi{10.1016/0019-1035(90)90181-8}.
\bibitem[{{Koseki} et~al.(1998){Koseki}, {Teranishi}, {Shiba} and
  {Sekiguchi}}]{Koseki1998}
\bibinfo{author}{{Koseki}, M.}, \bibinfo{author}{{Teranishi}, K.},
  \bibinfo{author}{{Shiba}, J.}, \bibinfo{author}{{Sekiguchi}, Y.},
  \bibinfo{year}{1998}.
\newblock \bibinfo{title}{{Giacobinids Returned in the Japanese Sky: Video and
  Photographic Observations}}.
\newblock \bibinfo{journal}{WGN, Journal of the International Meteor
  Organization} \bibinfo{volume}{26}, \bibinfo{pages}{260--262}.
\bibitem[{{Koten} et~al.(2007){Koten}, {Borovi{\v c}ka}, {Spurn{\'y}} and {{\v
  S}tork}}]{Koten2007}
\bibinfo{author}{{Koten}, P.}, \bibinfo{author}{{Borovi{\v c}ka}, J.},
  \bibinfo{author}{{Spurn{\'y}}, P.}, \bibinfo{author}{{{\v S}tork}, R.},
  \bibinfo{year}{2007}.
\newblock \bibinfo{title}{{Optical observations of enhanced activity of the
  2005 Draconid meteor shower}}.
\newblock \bibinfo{journal}{Astronomy and Astrophysics} \bibinfo{volume}{466},
  \bibinfo{pages}{729--735}.
\newblock \DOIprefix\doi{10.1051/0004-6361:20066838}.
\bibitem[{{Koten} et~al.(2014){Koten}, {Vaubaillon}, {T{\'o}th}, {Margonis} and
  {{\v D}uri{\v s}}}]{Koten2014}
\bibinfo{author}{{Koten}, P.}, \bibinfo{author}{{Vaubaillon}, J.},
  \bibinfo{author}{{T{\'o}th}, J.}, \bibinfo{author}{{Margonis}, A.},
  \bibinfo{author}{{{\v D}uri{\v s}}, F.}, \bibinfo{year}{2014}.
\newblock \bibinfo{title}{{Three Peaks of 2011 Draconid Activity Including that
  Connected with Pre-1900 Material}}.
\newblock \bibinfo{journal}{Earth Moon and Planets} \bibinfo{volume}{112},
  \bibinfo{pages}{15--31}.
\newblock \DOIprefix\doi{10.1007/s11038-014-9435-9}.
\bibitem[{{Kres{\'a}k}(1993)}]{Kresak1993}
\bibinfo{author}{{Kres{\'a}k}, {\v L}.}, \bibinfo{year}{1993}.
\newblock \bibinfo{title}{{Meteor storms (Invited)}}, in:
  \bibinfo{editor}{{Stohl}, J.}, \bibinfo{editor}{{Williams}, I.P.} (Eds.),
  \bibinfo{booktitle}{Meteoroids and their Parent Bodies}, p.
  \bibinfo{pages}{147}.
\bibitem[{{Kresak} and {Slancikova}(1975)}]{Kresak1975}
\bibinfo{author}{{Kresak}, L.}, \bibinfo{author}{{Slancikova}, J.},
  \bibinfo{year}{1975}.
\newblock \bibinfo{title}{{On the structure of the Giacobinid meteor shower}}.
\newblock \bibinfo{journal}{Bulletin of the Astronomical Institutes of
  Czechoslovakia} \bibinfo{volume}{26}, \bibinfo{pages}{327--342}.
\bibitem[{Kr{\'{o}}likowska et~al.(2001)Kr{\'{o}}likowska, Sitarski and
  Szutowicz}]{Krolikowska2001}
\bibinfo{author}{Kr{\'{o}}likowska, M.}, \bibinfo{author}{Sitarski, G.},
  \bibinfo{author}{Szutowicz, S.}, \bibinfo{year}{2001}.
\newblock \bibinfo{title}{{Forced precession models for six erratic comets M.}}
\newblock \bibinfo{journal}{A{\&}A} \DOIprefix\doi{10.1051/0004-6361},
  \href{http://arxiv.org/abs/arXiv:0709.4560v1}{\tt arXiv:arXiv:0709.4560v1}.
\bibitem[{{Lamy} et~al.(2004){Lamy}, {Toth}, {Fernandez} and
  {Weaver}}]{Lamy2004}
\bibinfo{author}{{Lamy}, P.L.}, \bibinfo{author}{{Toth}, I.},
  \bibinfo{author}{{Fernandez}, Y.R.}, \bibinfo{author}{{Weaver}, H.A.},
  \bibinfo{year}{2004}.
\newblock \bibinfo{title}{{The sizes, shapes, albedos, and colors of cometary
  nuclei}}.
\newblock pp. \bibinfo{pages}{223--264}.
\bibitem[{Landaberry et~al.(1991)Landaberry, Singh and
  Pacheco}]{Landaberry1991}
\bibinfo{author}{Landaberry, S.J.C.}, \bibinfo{author}{Singh, P.D.},
  \bibinfo{author}{Pacheco, J.A.d.F.}, \bibinfo{year}{1991}.
\newblock \bibinfo{title}{{Ground-based observations of comets Giacobini-Zinner
  (1984e) and Hartley-Good (1985I)}}.
\newblock \bibinfo{journal}{Astronomy and Astrophysics} .
\bibitem[{{Langbroek}(1999)}]{Langbroek1999}
\bibinfo{author}{{Langbroek}, M.}, \bibinfo{year}{1999}.
\newblock \bibinfo{title}{{The 1999 Draconids from the Netherlands and the
  Draconids of 1953.}}
\newblock \bibinfo{journal}{WGN, Journal of the International Meteor
  Organization} \bibinfo{volume}{27}, \bibinfo{pages}{335--338}.
\bibitem[{Lara et~al.(2003)Lara, Licandro, Oscoz and Motta}]{Lara2003}
\bibinfo{author}{Lara, L.M.}, \bibinfo{author}{Licandro, J.},
  \bibinfo{author}{Oscoz, A.}, \bibinfo{author}{Motta, V.},
  \bibinfo{year}{2003}.
\newblock \bibinfo{title}{{Behaviour of Comet 21P/Giacobini-Zinner during the
  1998 perihelion}}.
\newblock \bibinfo{journal}{A{\&}A} .
\bibitem[{{Leibowitz} and {Brosch}(1986)}]{Leibowitz1986}
\bibinfo{author}{{Leibowitz}, E.M.}, \bibinfo{author}{{Brosch}, N.},
  \bibinfo{year}{1986}.
\newblock \bibinfo{title}{{Periodic photometric variations in the near-nucleus
  zone of P/Giacobini-Zinner}}.
\newblock \bibinfo{journal}{Icarus} \bibinfo{volume}{68},
  \bibinfo{pages}{430--441}.
\newblock \DOIprefix\doi{10.1016/0019-1035(86)90049-7}.
\bibitem[{{Lindblad}(1987)}]{Lindblad1987}
\bibinfo{author}{{Lindblad}, B.A.}, \bibinfo{year}{1987}.
\newblock \bibinfo{title}{{The 1985 Return of the Giacobinid Meteor Stream}}.
\newblock \bibinfo{journal}{Astronomy and Astrophysics} \bibinfo{volume}{187},
  \bibinfo{pages}{928}.
\bibitem[{{Lovell} et~al.(1947){Lovell}, {Banwell} and {Clegg}}]{Lovell1947}
\bibinfo{author}{{Lovell}, A.C.B.}, \bibinfo{author}{{Banwell}, C.J.},
  \bibinfo{author}{{Clegg}, J.A.}, \bibinfo{year}{1947}.
\newblock \bibinfo{title}{{Radio echo observations of the Giacobinids meteors,
  1946}}.
\newblock \bibinfo{journal}{Monthly Notices of the RAS} \bibinfo{volume}{107},
  \bibinfo{pages}{164}.
\newblock \DOIprefix\doi{10.1093/mnras/107.2.164}.
\bibitem[{{Marsden}(1972)}]{IAU1972}
\bibinfo{author}{{Marsden}, B.G.}, \bibinfo{year}{1972}.
\newblock \bibinfo{title}{{IAU Circular No. 2451}}.
\bibitem[{{Marsden} and {Sekanina}(1971)}]{Marsden1971}
\bibinfo{author}{{Marsden}, B.G.}, \bibinfo{author}{{Sekanina}, Z.},
  \bibinfo{year}{1971}.
\newblock \bibinfo{title}{{Comets and Nongravitational Forces. IV}}.
\newblock \bibinfo{journal}{Astronomical Journal} \bibinfo{volume}{76},
  \bibinfo{pages}{1135}.
\newblock \DOIprefix\doi{10.1086/111232}.
\bibitem[{{Maslov}(2011)}]{Maslov2011}
\bibinfo{author}{{Maslov}, M.}, \bibinfo{year}{2011}.
\newblock \bibinfo{title}{{Future Draconid outbursts (2011 - 2100)}}.
\newblock \bibinfo{journal}{WGN, Journal of the International Meteor
  Organization} \bibinfo{volume}{39}, \bibinfo{pages}{64--67}.
\bibitem[{{Mason}(1986)}]{Mason1986}
\bibinfo{author}{{Mason}, J.W.}, \bibinfo{year}{1986}.
\newblock \bibinfo{title}{{Giacobinid meteor stream activity in October 1985}},
  in: \bibinfo{editor}{{Battrick}, B.}, \bibinfo{editor}{{Rolfe}, E.J.},
  \bibinfo{editor}{{Reinhard}, R.} (Eds.), \bibinfo{booktitle}{ESLAB Symposium
  on the Exploration of Halley's Comet}.
\bibitem[{{McBeath}(2012)}]{McBeath2012}
\bibinfo{author}{{McBeath}, A.}, \bibinfo{year}{2012}.
\newblock \bibinfo{title}{{SPA Meteor Section Results: Radio Draconids 2011}}.
\newblock \bibinfo{journal}{WGN, Journal of the International Meteor
  Organization} \bibinfo{volume}{40}, \bibinfo{pages}{126--128}.
\bibitem[{{McDonnell} et~al.(1987){McDonnell}, {Evans}, {Evans}, {Alexander},
  {Burton}, {Firth}, {Bussoletti}, {Grard}, {Hanner} and
  {Sekanina}}]{McDonnell1987}
\bibinfo{author}{{McDonnell}, J.A.M.}, \bibinfo{author}{{Evans}, G.C.},
  \bibinfo{author}{{Evans}, S.T.}, \bibinfo{author}{{Alexander}, W.M.},
  \bibinfo{author}{{Burton}, W.M.}, \bibinfo{author}{{Firth}, J.G.},
  \bibinfo{author}{{Bussoletti}, E.}, \bibinfo{author}{{Grard}, R.J.L.},
  \bibinfo{author}{{Hanner}, M.S.}, \bibinfo{author}{{Sekanina}, Z.},
  \bibinfo{year}{1987}.
\newblock \bibinfo{title}{{The dust distribution within the inner coma of comet
  P/Halley 1982i - Encounter by Giotto's impact detectors}}.
\newblock \bibinfo{journal}{\aap} \bibinfo{volume}{187},
  \bibinfo{pages}{719--741}.
\bibitem[{McFadden et~al.(1987)McFadden, A'Hearn, Feldman, Bohnhardt, Rahe and
  Festou}]{McFadden1987}
\bibinfo{author}{McFadden, L.A.}, \bibinfo{author}{A'Hearn, M.F.},
  \bibinfo{author}{Feldman, P.D.}, \bibinfo{author}{Bohnhardt, H.},
  \bibinfo{author}{Rahe, J.}, \bibinfo{author}{Festou, M.C.},
  \bibinfo{year}{1987}.
\newblock \bibinfo{title}{{Ultraviolet Spectrophotometry of Comet
  Giacobini-Zinner during the ICE Encounter}}.
\newblock \bibinfo{journal}{Icarus} \bibinfo{volume}{337}.
\bibitem[{{McIntosh}(1972)}]{McIntosh1972}
\bibinfo{author}{{McIntosh}, B.A.}, \bibinfo{year}{1972}.
\newblock \bibinfo{title}{{Giacobinid Meteor Shower: Prospect for 1972}}.
\newblock \bibinfo{journal}{Journal of the RAS of Canada} \bibinfo{volume}{66},
  \bibinfo{pages}{149}.
\bibitem[{{Millman}(1973)}]{Millman1973}
\bibinfo{author}{{Millman}, P.M.}, \bibinfo{year}{1973}.
\newblock \bibinfo{title}{{Meteor News: Airborne Observations of the 1972
  Giacobinids}}.
\newblock \bibinfo{journal}{Journal of the Royal Astronomical Society of
  Canada} \bibinfo{volume}{67}, \bibinfo{pages}{35}.
\bibitem[{{Molau} and {Barentsen}(2014)}]{Molau2014}
\bibinfo{author}{{Molau}, S.}, \bibinfo{author}{{Barentsen}, G.},
  \bibinfo{year}{2014}.
\newblock \bibinfo{title}{{Real-Time Flux Density Measurements of the 2011
  Draconid Meteor Outburst}}.
\newblock \bibinfo{journal}{Earth Moon and Planets} \bibinfo{volume}{112},
  \bibinfo{pages}{1--5}.
\newblock \DOIprefix\doi{10.1007/s11038-013-9425-3},
  \href{http://arxiv.org/abs/1312.3605}{\tt arXiv:1312.3605}.
\bibitem[{Moser(2018)}]{Moser2018}
\bibinfo{author}{Moser, D.}, \bibinfo{year}{2018}.
\newblock \bibinfo{title}{{MSFC Meteoroid Stream Model Results, 2019}}.
\newblock \bibinfo{type}{Technical Report}. JPID-FY18-02018.
\bibitem[{Moser and Cooke(2004)}]{Moser2004}
\bibinfo{author}{Moser, D.E.}, \bibinfo{author}{Cooke, W.J.},
  \bibinfo{year}{2004}.
\newblock \bibinfo{title}{Msfc stream model preliminary results: Modeling
  recent leonid and perseid encounters}.
\newblock \bibinfo{journal}{Earth, Moon, and Planets} \bibinfo{volume}{95},
  \bibinfo{pages}{141--153}.
\newblock \URLprefix \url{https://doi.org/10.1007/s11038-005-3185-7},
  \DOIprefix\doi{10.1007/s11038-005-3185-7}.
\bibitem[{{Moser} and {Cooke}(2008)}]{Moser2008}
\bibinfo{author}{{Moser}, D.E.}, \bibinfo{author}{{Cooke}, W.J.},
  \bibinfo{year}{2008}.
\newblock \bibinfo{title}{{Updates to the MSFC Meteoroid Stream Model}}.
\newblock \bibinfo{journal}{Earth Moon and Planets} \bibinfo{volume}{102},
  \bibinfo{pages}{285--291}.
\newblock \DOIprefix\doi{10.1007/s11038-007-9159-1}.
\bibitem[{Newburn and Spinrad(1985)}]{Newburn1985}
\bibinfo{author}{Newburn, R.L.}, \bibinfo{author}{Spinrad, H.},
  \bibinfo{year}{1985}.
\newblock \bibinfo{title}{{Spectrophotometry of seventeen comets. II. The
  continuum}}.
\newblock \bibinfo{journal}{The Astronomical Journal} \bibinfo{volume}{90},
  \bibinfo{pages}{2591--2608}.
\bibitem[{{Newburn} and {Spinrad}(1989)}]{Newburn1989}
\bibinfo{author}{{Newburn}, R.L.}, \bibinfo{author}{{Spinrad}, H.},
  \bibinfo{year}{1989}.
\newblock \bibinfo{title}{{Spectrophotometry of 25 comets - Post-Halley updates
  for 17 comets plus new observations for eight additional comets}}.
\newblock \bibinfo{journal}{\aj} \bibinfo{volume}{97},
  \bibinfo{pages}{552--569}.
\newblock \DOIprefix\doi{10.1086/115005}.
\bibitem[{{Olivier}(1946)}]{Olivier1946}
\bibinfo{author}{{Olivier}, C.P.}, \bibinfo{year}{1946}.
\newblock \bibinfo{title}{{Meteor notes from the American Meteor Society}}.
\newblock \bibinfo{journal}{Popular Astronomy} \bibinfo{volume}{54},
  \bibinfo{pages}{475}.
\bibitem[{{Pittichov{\'a}} et~al.(2008){Pittichov{\'a}}, {Woodward}, {Kelley}
  and {Reach}}]{Pittichova2008}
\bibinfo{author}{{Pittichov{\'a}}, J.}, \bibinfo{author}{{Woodward}, C.E.},
  \bibinfo{author}{{Kelley}, M.S.}, \bibinfo{author}{{Reach}, W.T.},
  \bibinfo{year}{2008}.
\newblock \bibinfo{title}{{Ground-Based Optical and Spitzer Infrared Imaging
  Observations of Comet 21P/GIACOBINI-ZINNER}}.
\newblock \bibinfo{journal}{Astronomical Journal} \bibinfo{volume}{136},
  \bibinfo{pages}{1127--1136}.
\newblock \DOIprefix\doi{10.1088/0004-6256/136/3/1127},
  \href{http://arxiv.org/abs/0806.3582}{\tt arXiv:0806.3582}.
\bibitem[{{Plavec}(1957)}]{Plavec1957}
\bibinfo{author}{{Plavec}, M.}, \bibinfo{year}{1957}.
\newblock \bibinfo{title}{{Vznik a rana vyvojova stadia meteorickyck roju. On
  the origin and early stages of the meteor streams.}}
\newblock \bibinfo{journal}{Publications of the Astronomical Institute of the
  Czechoslovak Academy of Sciences} \bibinfo{volume}{30}.
\bibitem[{{Pokorn{\'y}} and {Brown}(2016)}]{Pokorny2016}
\bibinfo{author}{{Pokorn{\'y}}, P.}, \bibinfo{author}{{Brown}, P.G.},
  \bibinfo{year}{2016}.
\newblock \bibinfo{title}{{A reproducible method to determine the meteoroid
  mass index}}.
\newblock \bibinfo{journal}{\aap} \bibinfo{volume}{592}, \bibinfo{pages}{A150}.
\newblock \DOIprefix\doi{10.1051/0004-6361/201628134},
  \href{http://arxiv.org/abs/1605.04437}{\tt arXiv:1605.04437}.
\bibitem[{{Rendtel} et~al.(2017){Rendtel}, {Ogawa} and
  {Sugimoto}}]{Rendtel2017}
\bibinfo{author}{{Rendtel}, J.}, \bibinfo{author}{{Ogawa}, H.},
  \bibinfo{author}{{Sugimoto}, H.}, \bibinfo{year}{2017}.
\newblock \bibinfo{title}{{Meteor showers 2016: review of predictions and
  observations}}.
\newblock \bibinfo{journal}{WGN, Journal of the International Meteor
  Organization} \bibinfo{volume}{45}, \bibinfo{pages}{49--55}.
\bibitem[{{Reznikov}(1993)}]{Reznikov1993}
\bibinfo{author}{{Reznikov}, E.A.}, \bibinfo{year}{1993}.
\newblock \bibinfo{title}{{The Giacobini-Zinner comet and Giacobinid meteor
  stream}}.
\newblock \bibinfo{journal}{Trudy Kazan. Gor. Astron. Obs.}
  \bibinfo{volume}{53}, \bibinfo{pages}{80--101}.
\bibitem[{{Sato}(2003)}]{Sato2003}
\bibinfo{author}{{Sato}, M.}, \bibinfo{year}{2003}.
\newblock \bibinfo{title}{{An investigation into the 1998 and 1999 Giacobinids
  by meteoroid trajectory modeling}}.
\newblock \bibinfo{journal}{WGN, Journal of the International Meteor
  Organization} \bibinfo{volume}{31}, \bibinfo{pages}{59--63}.
\bibitem[{Schleicher et~al.(1987)Schleicher, Millis and Birch}]{Schleicher1987}
\bibinfo{author}{Schleicher, D.G.}, \bibinfo{author}{Millis, R.L.},
  \bibinfo{author}{Birch, P.V.}, \bibinfo{year}{1987}.
\newblock \bibinfo{title}{{Photometric observations of comet
  P/Giacobini-Zinner}}.
\newblock \bibinfo{journal}{A{\&}A} .
\bibitem[{Sekanina(1985)}]{Sekanina1985}
\bibinfo{author}{Sekanina, Z.}, \bibinfo{year}{1985}.
\newblock \bibinfo{title}{{Precession model for the nucleus of periodic comet
  Giacobini/Zinner}}.
\newblock \bibinfo{journal}{The Astronomical Journal} \bibinfo{volume}{90}.
\bibitem[{{Sekanina}(1993)}]{Sekanina1993}
\bibinfo{author}{{Sekanina}, Z.}, \bibinfo{year}{1993}.
\newblock \bibinfo{title}{{Effects of discrete-source outgassing on motions of
  periodic comets and discontinuous orbital anomalies}}.
\newblock \bibinfo{journal}{Astronomical Journal} \bibinfo{volume}{105},
  \bibinfo{pages}{702--735}.
\newblock \DOIprefix\doi{10.1086/116468}.
\bibitem[{{Sidorov} et~al.(1994){Sidorov}, {Stepanov} and
  {Sulejmanov}}]{Sidorov1994}
\bibinfo{author}{{Sidorov}, V.V.}, \bibinfo{author}{{Stepanov}, A.M.},
  \bibinfo{author}{{Sulejmanov}, N.I.}, \bibinfo{year}{1994}.
\newblock \bibinfo{title}{{Observations of the Draconid meteor stream in Kazan
  in 1985.}}
\newblock \bibinfo{journal}{Solar System Research} \bibinfo{volume}{28},
  \bibinfo{pages}{190--193}.
\bibitem[{{Sigismondi}(2011)}]{Sigismondi2011}
\bibinfo{author}{{Sigismondi}, C.}, \bibinfo{year}{2011}.
\newblock \bibinfo{title}{{Airborne observation of 2011 Draconids meteor
  outburst: the Italian mission}}.
\newblock \bibinfo{journal}{ArXiv e-prints}
  \href{http://arxiv.org/abs/1112.4873}{\tt arXiv:1112.4873}.
\bibitem[{{Simek}(1986)}]{Simek1986}
\bibinfo{author}{{Simek}, M.}, \bibinfo{year}{1986}.
\newblock \bibinfo{title}{{Radar observation of the Giacobinid meteor shower
  1985}}.
\newblock \bibinfo{journal}{Bulletin of the Astronomical Institutes of
  Czechoslovakia} \bibinfo{volume}{37}, \bibinfo{pages}{246--249}.
\bibitem[{{Simek}(1994)}]{Simek1994}
\bibinfo{author}{{Simek}, M.}, \bibinfo{year}{1994}.
\newblock \bibinfo{title}{{Fine structure of the 1985 Giacobinids}}.
\newblock \bibinfo{journal}{Astronomy and Astrophysics} \bibinfo{volume}{284},
  \bibinfo{pages}{276--280}.
\bibitem[{Singh et~al.(1997)Singh, Huebner, Costa, Landaberry and
  Pacheco}]{Singh1997}
\bibinfo{author}{Singh, P.D.}, \bibinfo{author}{Huebner, W.F.},
  \bibinfo{author}{Costa, R.D.D.}, \bibinfo{author}{Landaberry, S.J.C.},
  \bibinfo{author}{Pacheco, J.A.d.F.}, \bibinfo{year}{1997}.
\newblock \bibinfo{title}{{Gas and dust release rates and color of dust in
  comets P/Halley (1986 III), P/G iacobini-Zinner (1985 XIII), and
  P/Hartley-Good (1985 XVII)}}.
\newblock \bibinfo{journal}{Planet. Space Sci.} \bibinfo{volume}{45}.
\bibitem[{{Southworth} and {Hawkins}(1963)}]{Southworth1963}
\bibinfo{author}{{Southworth}, R.B.}, \bibinfo{author}{{Hawkins}, G.S.},
  \bibinfo{year}{1963}.
\newblock \bibinfo{title}{{Statistics of meteor streams}}.
\newblock \bibinfo{journal}{Smithsonian Contributions to Astrophysics}
  \bibinfo{volume}{7}, \bibinfo{pages}{261}.
\bibitem[{{Tody}(1986)}]{Tody1986}
\bibinfo{author}{{Tody}, D.}, \bibinfo{year}{1986}.
\newblock \bibinfo{title}{{The IRAF Data Reduction and Analysis System}}, in:
  \bibinfo{editor}{{Crawford}, D.L.} (Ed.), \bibinfo{booktitle}{Instrumentation
  in astronomy VI}, p. \bibinfo{pages}{733}.
\newblock \DOIprefix\doi{10.1117/12.968154}.
\bibitem[{{Toth} et~al.(2012){Toth}, {Koukal}, {Zoladek}, {Wisniewski},
  {Gajdos}, {Zanotti}, {Valeri}, {De Maria}, {Popek}, {Gorkova}, {Vilagi},
  {Kornos}, {Kalmancok} and {Zigo}}]{Toth2012}
\bibinfo{author}{{Toth}, J. an~{Piffl}, R.}, \bibinfo{author}{{Koukal}, J.},
  \bibinfo{author}{{Zoladek}, P.}, \bibinfo{author}{{Wisniewski}, M.},
  \bibinfo{author}{{Gajdos}, S.}, \bibinfo{author}{{Zanotti}, F.},
  \bibinfo{author}{{Valeri}, D.}, \bibinfo{author}{{De Maria}, P.},
  \bibinfo{author}{{Popek}, M.}, \bibinfo{author}{{Gorkova}, S.},
  \bibinfo{author}{{Vilagi}, J.}, \bibinfo{author}{{Kornos}, L.},
  \bibinfo{author}{{Kalmancok}, D.}, \bibinfo{author}{{Zigo}, P.},
  \bibinfo{year}{2012}.
\newblock \bibinfo{title}{{Video observation of Draconids 2011 from Italy}}.
\newblock \bibinfo{journal}{WGN, Journal of the International Meteor
  Organization} \bibinfo{volume}{40}, \bibinfo{pages}{117--121}.
\bibitem[{{Trigo-Rodr{\'{\i}}guez} et~al.(2013){Trigo-Rodr{\'{\i}}guez},
  {Madiedo}, {Williams}, {Dergham}, {Cort{\'e}s}, {Castro-Tirado}, {Ortiz},
  {Zamorano}, {Oca{\~n}a}, {Izquierdo}, {S{\'a}nchez de Miguel},
  {Alonso-Azc{\'a}rate}, {Rodr{\'{\i}}guez}, {Tapia}, {Pujols}, {Lacruz},
  {Pruneda}, {Oliva}, {Pastor Erades} and {Francisco Mar{\'{\i}}n}}]{Trigo2013}
\bibinfo{author}{{Trigo-Rodr{\'{\i}}guez}, J.M.}, \bibinfo{author}{{Madiedo},
  J.M.}, \bibinfo{author}{{Williams}, I.P.}, \bibinfo{author}{{Dergham}, J.},
  \bibinfo{author}{{Cort{\'e}s}, J.}, \bibinfo{author}{{Castro-Tirado}, A.J.},
  \bibinfo{author}{{Ortiz}, J.L.}, \bibinfo{author}{{Zamorano}, J.},
  \bibinfo{author}{{Oca{\~n}a}, F.}, \bibinfo{author}{{Izquierdo}, J.},
  \bibinfo{author}{{S{\'a}nchez de Miguel}, A.},
  \bibinfo{author}{{Alonso-Azc{\'a}rate}, J.},
  \bibinfo{author}{{Rodr{\'{\i}}guez}, D.}, \bibinfo{author}{{Tapia}, M.},
  \bibinfo{author}{{Pujols}, P.}, \bibinfo{author}{{Lacruz}, J.},
  \bibinfo{author}{{Pruneda}, F.}, \bibinfo{author}{{Oliva}, A.},
  \bibinfo{author}{{Pastor Erades}, J.}, \bibinfo{author}{{Francisco
  Mar{\'{\i}}n}, A.}, \bibinfo{year}{2013}.
\newblock \bibinfo{title}{{The 2011 October Draconids outburst - I. Orbital
  elements, meteoroid fluxes and 21P/Giacobini-Zinner delivered mass to
  Earth}}.
\newblock \bibinfo{journal}{Monthly Notices of the RAS} \bibinfo{volume}{433},
  \bibinfo{pages}{560--570}.
\newblock \DOIprefix\doi{10.1093/mnras/stt749},
  \href{http://arxiv.org/abs/1304.7635}{\tt arXiv:1304.7635}.
\bibitem[{{{\v S}imek} and {Pecina}(1999)}]{Simek1999}
\bibinfo{author}{{{\v S}imek}, M.}, \bibinfo{author}{{Pecina}, P.},
  \bibinfo{year}{1999}.
\newblock \bibinfo{title}{{The Giacobinid meteor stream observed by radar in
  1998}}.
\newblock \bibinfo{journal}{Astronomy and Astrophysics} \bibinfo{volume}{343},
  \bibinfo{pages}{L94--L96}.
\bibitem[{{Vaubaillon} et~al.(2005){Vaubaillon}, {Colas} and
  {Jorda}}]{Vaubaillon2005}
\bibinfo{author}{{Vaubaillon}, J.}, \bibinfo{author}{{Colas}, F.},
  \bibinfo{author}{{Jorda}, L.}, \bibinfo{year}{2005}.
\newblock \bibinfo{title}{{A new method to predict meteor showers. I.
  Description of the model}}.
\newblock \bibinfo{journal}{Astronomy and Astrophysics} \bibinfo{volume}{439},
  \bibinfo{pages}{751--760}.
\newblock \DOIprefix\doi{10.1051/0004-6361:20041544}.
\bibitem[{{Vaubaillon} et~al.(2011){Vaubaillon}, {Watanabe}, {Sato}, {Horii}
  and {Koten}}]{Vaubaillon2011}
\bibinfo{author}{{Vaubaillon}, J.}, \bibinfo{author}{{Watanabe}, J.},
  \bibinfo{author}{{Sato}, M.}, \bibinfo{author}{{Horii}, S.},
  \bibinfo{author}{{Koten}, P.}, \bibinfo{year}{2011}.
\newblock \bibinfo{title}{{The coming 2011 Draconids meteor shower}}.
\newblock \bibinfo{journal}{WGN, Journal of the International Meteor
  Organization} \bibinfo{volume}{39}, \bibinfo{pages}{59--63}.
\bibitem[{{Vokrouhlick{\'y}} and {Farinella}(2000)}]{Vokrouhlicky2000}
\bibinfo{author}{{Vokrouhlick{\'y}}, D.}, \bibinfo{author}{{Farinella}, P.},
  \bibinfo{year}{2000}.
\newblock \bibinfo{title}{{Efficient delivery of meteorites to the Earth from a
  wide range of asteroid parent bodies}}.
\newblock \bibinfo{journal}{Nature} \bibinfo{volume}{407},
  \bibinfo{pages}{606--608}.
\newblock \DOIprefix\doi{10.1038/35036528}.
\bibitem[{{Watanabe} et~al.(1999){Watanabe}, {Abe}, {Takanashi}, {Hashimoto},
  {Iiyama}, {Ishibashi}, {Morishige} and {Yokogawa}}]{Watanabe1999}
\bibinfo{author}{{Watanabe}, J.i.}, \bibinfo{author}{{Abe}, S.},
  \bibinfo{author}{{Takanashi}, M.}, \bibinfo{author}{{Hashimoto}, T.},
  \bibinfo{author}{{Iiyama}, O.}, \bibinfo{author}{{Ishibashi}, Y.},
  \bibinfo{author}{{Morishige}, K.}, \bibinfo{author}{{Yokogawa}, S.},
  \bibinfo{year}{1999}.
\newblock \bibinfo{title}{{HD TV observation of the strong activity of the
  Giacobinid Meteor Shower in 1998}}.
\newblock \bibinfo{journal}{Geophysical Research Letters} \bibinfo{volume}{26},
  \bibinfo{pages}{1117--1120}.
\newblock \DOIprefix\doi{10.1029/1999GL900195}.
\bibitem[{{Watanabe} and {Sato}(2008)}]{Watanabe2008}
\bibinfo{author}{{Watanabe}, J.I.}, \bibinfo{author}{{Sato}, M.},
  \bibinfo{year}{2008}.
\newblock \bibinfo{title}{{Activities of Parent Comets and Related Meteor
  Showers}}.
\newblock \bibinfo{journal}{Earth Moon and Planets} \bibinfo{volume}{102},
  \bibinfo{pages}{111--116}.
\newblock \DOIprefix\doi{10.1007/s11038-007-9193-z}.
\bibitem[{{Watson}(1934)}]{Watson1934}
\bibinfo{author}{{Watson}, Jr., F.}, \bibinfo{year}{1934}.
\newblock \bibinfo{title}{{Luminosity Function of the Giacobinid Meteors}}.
\newblock \bibinfo{journal}{Harvard College Observatory Bulletin}
  \bibinfo{volume}{895}, \bibinfo{pages}{9--16}.
\bibitem[{{Whipple}(1951)}]{Whipple1951}
\bibinfo{author}{{Whipple}, F.L.}, \bibinfo{year}{1951}.
\newblock \bibinfo{title}{{A Comet Model. II. Physical Relations for Comets and
  Meteors.}}
\newblock \bibinfo{journal}{Astrophysical Journal} \bibinfo{volume}{113},
  \bibinfo{pages}{464}.
\newblock \DOIprefix\doi{10.1086/145416}.
\bibitem[{Wu and Williams(1995)}]{Wu1995}
\bibinfo{author}{Wu, Z.}, \bibinfo{author}{Williams, I.P.},
  \bibinfo{year}{1995}.
\newblock \bibinfo{title}{{P/Giacobini-Zinner}}.
\newblock \bibinfo{journal}{Planetary and Space Science} \bibinfo{volume}{43},
  \bibinfo{pages}{723--731}.
\bibitem[{{Ye} et~al.(2014){Ye}, {Wiegert}, {Brown}, {Campbell-Brown} and
  {Weryk}}]{Ye2014}
\bibinfo{author}{{Ye}, Q.}, \bibinfo{author}{{Wiegert}, P.A.},
  \bibinfo{author}{{Brown}, P.G.}, \bibinfo{author}{{Campbell-Brown}, M.D.},
  \bibinfo{author}{{Weryk}, R.J.}, \bibinfo{year}{2014}.
\newblock \bibinfo{title}{{The unexpected 2012 Draconid meteor storm}}.
\newblock \bibinfo{journal}{Monthly Notices of the Royal Astronomical Society}
  \bibinfo{volume}{437}, \bibinfo{pages}{3812--3823}.
\newblock \DOIprefix\doi{10.1093/mnras/stt2178},
  \href{http://arxiv.org/abs/1311.1733}{\tt arXiv:1311.1733}.
\bibitem[{{Yeomans}(1971)}]{Yeomans1971}
\bibinfo{author}{{Yeomans}, D.K.}, \bibinfo{year}{1971}.
\newblock \bibinfo{title}{{Nongravitational Forces Affecting.- the Motions of
  Periodic Comets Giacobini-Ziner and Borrelly}}.
\newblock \bibinfo{journal}{The Astronomical Journal} \bibinfo{volume}{76},
  \bibinfo{pages}{83}.
\newblock \DOIprefix\doi{10.1086/111089}.
\bibitem[{Yeomans(1972)}]{Yeomans1972}
\bibinfo{author}{Yeomans, D.K.}, \bibinfo{year}{1972}.
\newblock \bibinfo{title}{{Nongravitational forces and periodic comet
  Giacobini-Zinner}}, in: \bibinfo{booktitle}{The Motion, Evolution of Orbits,
  and Origin of Comets}.
\bibitem[{Yeomans(1986)}]{Yeomans1986}
\bibinfo{author}{Yeomans, D.K.}, \bibinfo{year}{1986}.
\newblock \bibinfo{title}{{Physical interpretations form the motions of comets
  Halley and Giacobini-Zinner}}.
\newblock \bibinfo{journal}{20th ESLAB Symposium} .
\bibitem[{Yeomans and Chodas(1989)}]{Yeomans1989}
\bibinfo{author}{Yeomans, D.K.}, \bibinfo{author}{Chodas, P.W.},
  \bibinfo{year}{1989}.
\newblock \bibinfo{title}{{An asymmetric outgassing model for cometary
  nongravitationnal accelerations}}.
\newblock \bibinfo{journal}{The Astronomical Journal} \bibinfo{volume}{98},
  \bibinfo{pages}{1083--1093}.

\end{thebibliography}

\clearpage
\appendix

  \section{Ejection velocity} \label{sec:Vej}

\begin{figure}[!ht]
    \centering
    \includegraphics[width=0.47\textwidth]{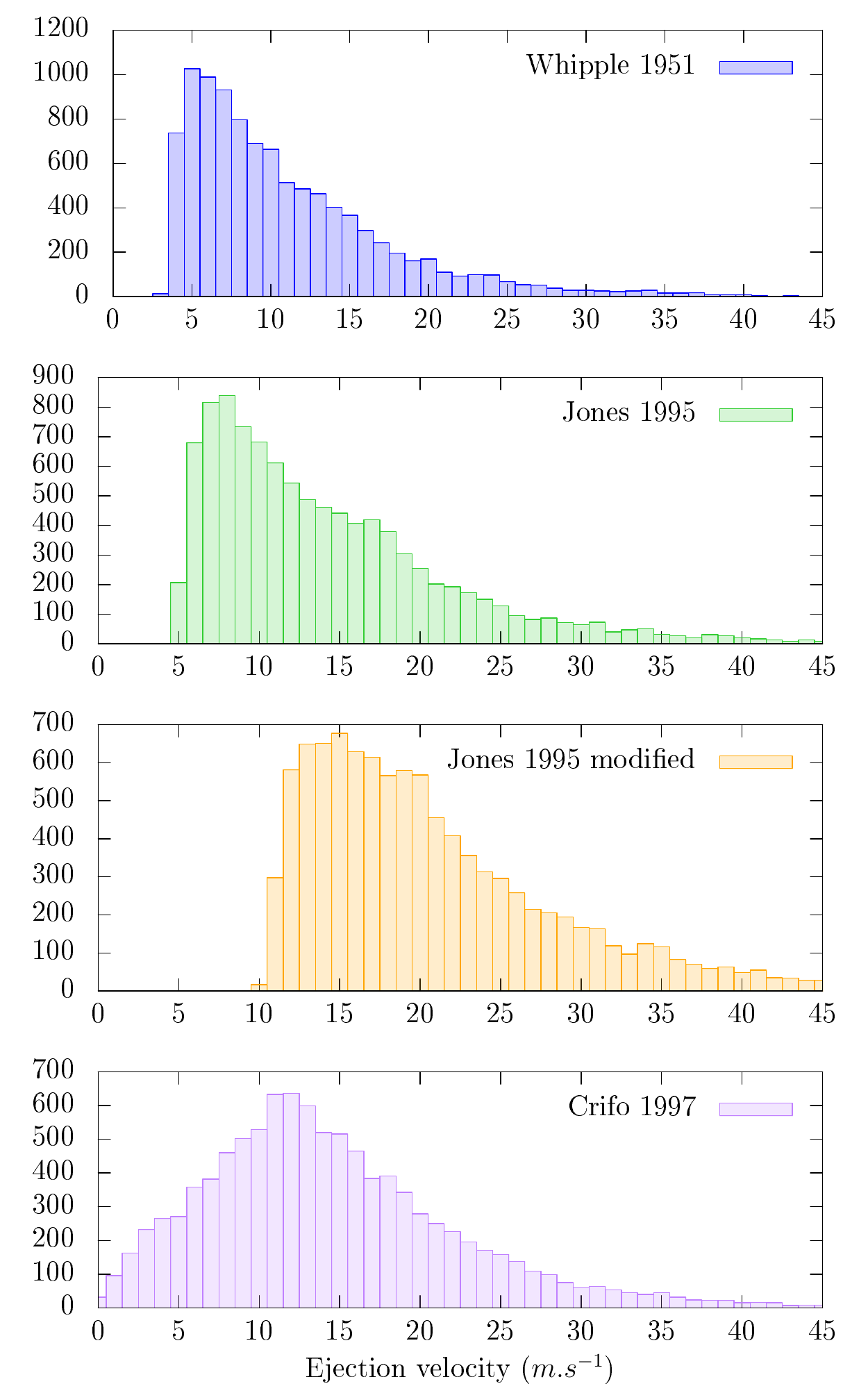}
    \caption{\label{fig:Vej} Ejection velocity distributions generated using the parameters of Table \ref{table:parameters} for different models.}
    \end{figure}

  In this work, we compared the performance of different ejection velocity models, using the parameters of Table \ref{table:parameters}. Four ejection models were specifically investigated: the original Whipple formula \citep{Whipple1951}, the Jones model \citep[][$r_h^{-1.038}$]{Jones1995}, the Jones model as modified by \cite[][$r_h^{-0.5}$]{Brown1998}, and the \cite{Crifo1997} model. The analytical expression of each ejection velocity implemented is given in Equation \ref{eq:Vej}, and the corresponding distributions are presented in Figure \ref{fig:Vej}. 
  
  \begin{equation} \label{eq:Vej}
  \begin{aligned}
   \text{Whipple 1951: }  V_{ej}&=\sqrt{r_c \left( \frac{43}{a \rho r_h^{2.25}}-\frac{8\pi\mathcal{G} \rho _c r_c}{3}   \right)}\\
    \text{Jones 1995: }  V_{ej}&=10.2 r_h^{-1.038}\rho^{-1/3}\sqrt{r_c}m^{-1/6} \\
   \text{Modified: } V_{ej}&=10.2 r_h^{-0.5}\rho^{-1/3}\sqrt{r_c}m^{-1/6}\\
   \text{Crifo 1997: } V_{ej}&=\frac{W}{1.2+0.72\sqrt{\frac{a}{a\star}}}  \hspace{0.4cm}\text{(cf. Eq. \ref{eq:VCfo})}\\ 
    \end{aligned}
  \end{equation}
  
Recent measurements performed by the Rosetta spacecraft around comet 67P/Churyumov-Gerasimenko suggest that the real ejection velocities of dust particles are much lower than the speeds usually considered by these models. As an example, most of the compact grains with masses of $[10^{-10},10^{-7}]$ kg detected by the Grain Impact Analyser and Dust Accumulator (GIADA) instrument displayed  velocities ranging from 0.3 to 12.2 m/s \citep{DellaCorte2015}. When looking at the distribution profiles drawn in Figure \ref{fig:Vej} we see that among the four formulas considered, the \cite{Crifo1997} model is the most efficient in ejecting particles with low speeds ($\in ]0,5]$ m/s). As in \cite{Vaubaillon2005}, we therefore decided to utilize this ejection model for our simulations.

\section{21P's $Af\rho$ evolution} \label{sec:comet_photometry}

In our weighting model (cf. \ref{sec:weights_eq}), we assume that the $Af\rho$ parameter defined by \cite{AHearn1984} directly reflects the number of dust particles ejected by comet 21P. The number of real meteoroids associated with each simulated particle therefore depends on the evolution of $Af\rho$ with the comet heliocentric distance. Quantities implied in this parameter are \citep{AHearn1984}:
\begin{enumerate}
 \item $A$: the albedo $A(\Phi)=4\pi A_B j(\Phi)$, which depends on the Bond albedo $A_B$ and the normalized phase function $j(\Phi)$ at angle $\Phi$ 
 \item $f$: the filling factor of the dust grains within the field of view, usually computed as the sum of the particles geometric cross sections divided by the area of the field of view
 \item $\rho$: the aperture radius used for the observation. This factor was added to make the $Af\rho$ parameter independent of the field of view observed. This assumption is asserted in the case of a simple radial-outflow cometary model with a brightness profile evolving as $1/\rho$.
\end{enumerate}

  While the brightness profile of comet 21P/Giacobini-Zinner may obey a $\rho^{-1}$ dependence  within a limited central region, the profile will eventually steepen for larger aperture radii \citep{Jewitt1987}. For this reason and because of the suspected variability of 21P's coma brightness on a daily time scale, one needs to be very cautious in comparing $Af\rho$ measurements of the comet performed at different epochs using different instruments and reduction techniques. Despite numerous observations of 21P performed since 1985 \citep[e.g.][]{Schleicher1987,Singh1997,Lara2003,Pittichova2008,Blaauw2014}, no general agreement in the $Af\rho$ estimates of the comet was reached for both the pre- and the post-perihelion branch of its orbit. To address this issue, an unprecedented observation campaign targeting 21P has been undertaken by the NASA Meteoroid Environment Office in May 2018, starting five months before the perihelion approach and continuing until 2019. 
  
  Telescopic observations of Comet 21P were taken on 65 nights spanning from 2018-05-05 through 2018-12-18 using publicly accessible telescopes owned and operated by iTelescope\footnote{\url{https://www.itelescope.net/}}. In particular, we utilized telescopes T11 in Mayhill New Mexico, T7 in Nerpio Spain, and T30/T31 at Siding Spring Observatory in Australia. These observations imaged the comet during its ingress phase at heliocentric distances ranging from 1.93 AU through perihelion (1.01 AU) and during its egress phase back out to 1.66 AU. Telescope T7 in Spain has a primary mirror diameter of $0.41 \thinspace \mathrm{m}$, while the other three telescopes all have primary mirrors $0.50 \thinspace \mathrm{m}$ in diameter. Ten images of the comet were taken each night using the Johnson-Cousins $R_{C}$ band filter, with exposure times ranging from $15$ seconds to $1$ minute based on the non-sidereal motion of the comet. Determinations of $Af\rho$ for each set of images were performed in a manner nearly identical to the one described in \cite{Hosek2013} and \cite{Blaauw2014}. Just like in these works, the $Af\rho$ measurements are performed within a $10^{\prime \prime} \times 10^{\prime \prime}$ square aperture centered on the brightest pixel associated with the comet emission. Further details on the observation campaign and the derived $Af\rho$ values are discussed in Ehlert et al. (in preparation). 
  
  A first analysis showed a noticeable asymmetry between the ascending activity branch (pre-perihelion measurements) and the descending branch (post-perihelion). When fitting each branch with a logarithmic function of the form $Af\rho(r_h)=K_1+K_2 r_h ^\gamma$, best fits were obtained for a logarithmic slope $\gamma=-5.5$ for the ascending branch and $\gamma=-15.5$ for the descending branch. However, no accurate fit of both branches was obtained when forcing the extrapolated maximal $Af\rho$ value to be identical for each fit. Because the weighting solution requires an analytical expression of $Af\rho=f(r_h)$ over all the comet orbit, we decided to adjust preliminary $Af\rho$ measurements with a standard double-exponential shape function of the form:

\begin{equation}\label{eq:afrho}
\left\{
  \begin{aligned}
  & Af\rho(X)=K_1+Af\rho(X_{max})*10^{-\gamma |X-X_{max}|} \\
  & \gamma=\gamma_1 \text{ if } X\ge X_{max} \text{ and } \gamma=\gamma_2 \text{ if } X\le X_{max} \\
    & X=(r_h-q)\frac{t(q)-t}{|t(q)-t|}\\
 \end{aligned}
 \right.
\end{equation}

\begin{figure}[!ht]
 \centering
\includegraphics[width=.47\textwidth]{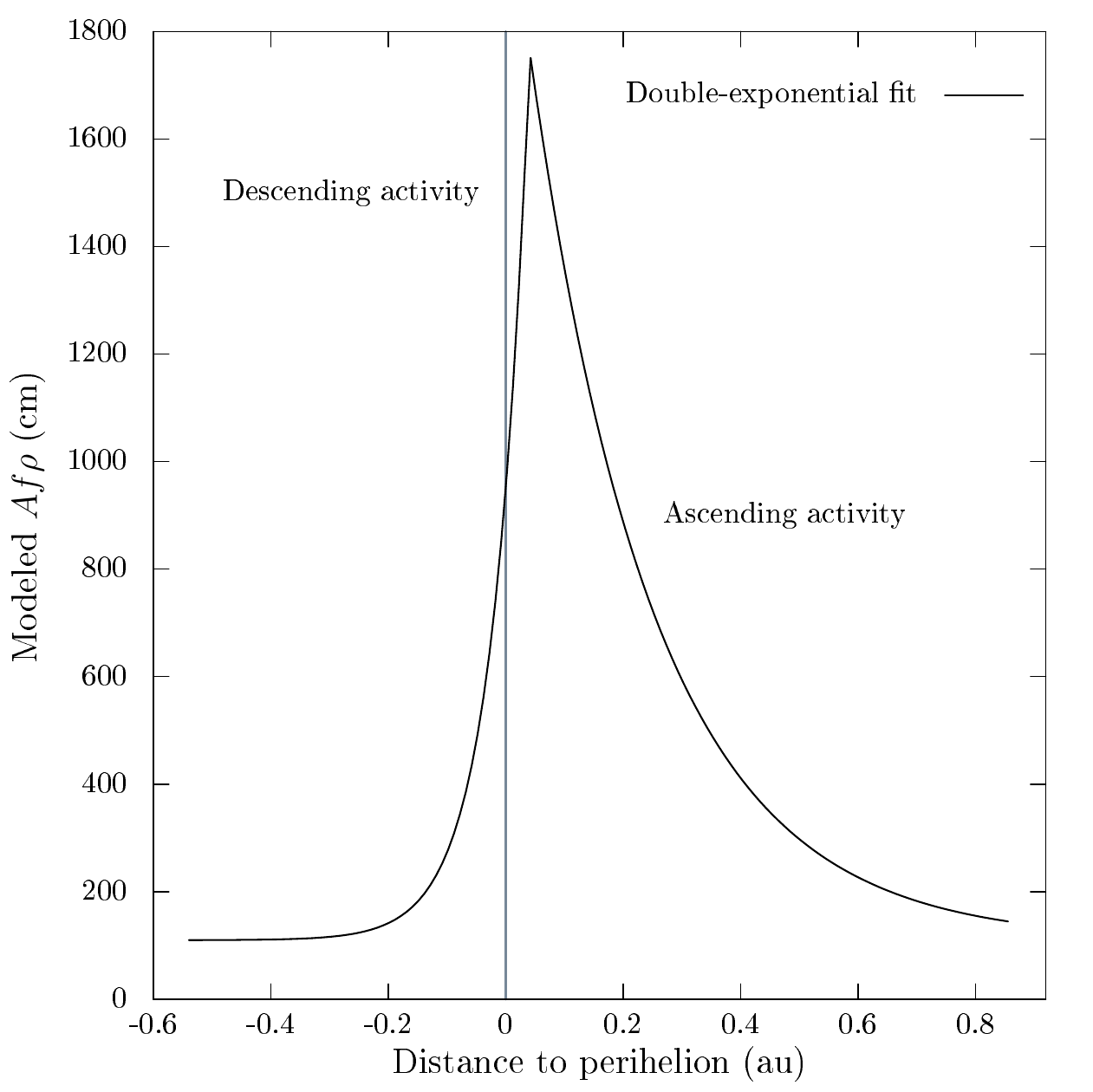}
\caption{ \label{fig:Afrho} Modeled $Af\rho$ dependence with the distance to perihelion for 21P's 2018 apparition. \zcorr{The abscissa represents } the heliocentric distance to perihelion, arbitrarily positive for pre-perihelion measurements and negative \zcorr{for post-perihelion dust production ($X=(r_h-q)\frac{t(q)-t}{|t(q)-t|}$}, with $t(q)$ the time at perihelion).}
\end{figure}

The best agreement with 21P's observations was reached  for $K_1\sim 100$, $Af\rho(X_{max})\sim1650$, $\gamma_1=2.0$,  $\gamma_2=7.0$ and $X_{max}=0.04$, and illustrated by Figure \ref{fig:Afrho}. Fit attempts using Gaussian, Lorentzian or Moffat functions were not more successful than the double-exponential shape model in reproducing the observations of the comet.

As already noticed by several authors \citep[e.g.][]{Lara2003,Blaauw2014}, preliminary observations performed in 2018 indicate a cometary activity not peaking exactly at the perihelion ($X_{max} \ne 0$).  It is assumed that the gas and dust productions peak before the perihelion since $\sim$1960, when the comet suffered dramatic changes in its non-gravitational coefficients \cite{Sekanina1985,Blaauw2014}. In 2018, the maximum dust production of the comet is observed a few days before the perihelion passage. This time shift varied for each apparition of the comet, reaching a value of about one month in 1985 \citep{McFadden1987,Schleicher1987}. Because of the lack of information regarding the asymmetric outgassing of the comet at each apparition, we however will consider that the maximum outgassing of the comet and its perihelion passage coincides for each apparition considered in our simulations ($X_{max}=0$). 

For the same reason, we are also forced to assume that the cometary dust emission with the heliocentric distance is similar for all orbits considered. $Af\rho(r_h)$ estimates for ancient locations of the comet therefore only rely on its orbital characteristics (e.g. perihelion distance and semi-major axis).  Even if these assumptions are likely erroneous given the variability of 21P, they provide a first approximation of the dust emission evolution for each apparition of the comet. 

\section{Weighting scheme} \label{sec:weights_eq}

This appendix summarizes the steps we take to determine the total number of meteoroids ejected by the comet at a given location. For more information, the reader is referred to \cite{Vaubaillon2005}. 
 
  \subsection{Gas production rate}
 
  We assume that the comet's outgassing is dominated by the sublimation of water ice. The global gas production rate $Q_{H_2O}(r_h)$ is a function of the heliocentric distance $r_h$. In this model, the water molecules and dust particles are ejected from the sunlit hemisphere of a spherical and homogeneous nucleus. The intensity of the gas production depends on the cosine of the angle $\theta$ between the ejection direction and the subsolar point. The effective sublimation rate for a unit area at the surface of the comet is then expressed as
  
     \begin{equation}
    Z_{\textrm{H}_2\textrm{O}}(r_h,\theta)= Q_{\textrm{H}_2\textrm{O}}(r_h)\frac{\cos\theta}{\pi R_c^2}
    \end{equation}
  
  where $R_c$ is the radius (in km) of the comet's nucleus. 
  
   \subsection{Dust production rate} 

   At a heliocentric distance $r_h$, the local dust sublimation rate $Z_g$ (in m$^{-2}$~s$^{-1}$) is related to the local ice sublimation rate by a variable dust-to-gas ratio $K(r_h)$: 
   
   \begin{equation}
     Z_g(r_h,\theta)=K(r_h)Z_{\textrm{H}_2\textrm{O}}(r_h,\theta)
   \end{equation}
   
   In contrast with \cite{Vaubaillon2005}, we do not assume that the function $K$ is independent of heliocentric distance. In order to correlate the local dust sublimation rate with observable measurements of the particles released by the comet, it is necessary to adopt a definite size distribution of the dust particles at ejection. Following \cite{Vaubaillon2005}, we assume that the size distribution $h$ of the particles follows a power law of index $u$ such that 
   
   \begin{equation}
      h(a)=\frac{N}{a^u}
   \end{equation}
where $a$ is the radius of the particles in m and $N$ is a normalization constant solution of the equation $\int_{a_1}^{a_2}h(a)da=1$. The differential local dust sublimation rate for a particle of size $a$ becomes 
    
   \begin{equation}\label{eq:localZ}
     Z_g(r_h,\theta,a)=K(r_h)Q_{\textrm{H}_2\textrm{O}}(r_h)\frac{\cos\theta}{\pi R_c^2}\frac{N}{a^u}
   \end{equation}

   The parameter $K(r_h)$ is determined from the $Af\rho$ measurement at the heliocentric distance $r_h$. Using Equation \ref{eq:localZ}, the density $n_g$ of particles of radius $a\in[a_1,a_2]$ around the nucleus is given by 
   
   \begin{equation}
    n_g(a,\theta,d,r_h)=\frac{Z_g(r_h,\theta,a)}{v_g(r_h,\theta,a)}\left(\frac{R_c}{d}\right)^2
   \end{equation}
   where $d$ is the distance to the nucleus (in km) and $v_g$ is  the dust terminal velocity in (m~s$^{-1}$). The radii $a_1$ \& $a_2$ represent respectively the minimal and maximal radius of a particle that can be ejected by the comet. The dust terminal velocity depends on the ejection model and the size and density of the particles considered. For the \cite{Crifo1997} ejection model, the terminal velocity of a particle with radius $a$, ejected at $r_h$ for an angle $\theta$ between the ejection direction and the subsolar point is 
    
    \begin{equation}\label{eq:VCfo}
     \begin{aligned}
      v_g(a,\theta,r_h)&=W\phi(\theta,a)=\frac{W}{\alpha+\beta\sqrt{\frac{a}{\cos(\theta)}}}\\
     \text{where \mbox{ }} W&=\sqrt{\frac{\gamma_{H_2O}+1}{\gamma_{H_2O}-1} \frac{\gamma_{H_2O} k_B T}{m}}
     \end{aligned}
    \end{equation}
    and 
    \begin{itemize}
     \item[-] $\gamma_{H_2O}=\frac{4}{3}$ is the ratio of the specific water heats,
     \item[-] $k_B$ is the Boltzmann constant,
     \item[-] $T$ is the gas temperature,
     \item[-] $m$ is the mass of gas molecules (18 amu),
     \item[-] $\alpha=1.2$,
     \item[-] $\beta=0.72/\sqrt{a_\star ^0}$, where $a_\star ^0$ is the critical radius of the particle below which the gas-particle interaction efficiency starts to decrease. For $a<< a_\star ^0$, $v_g(a,\theta,r_h)\simeq W$ \citep{Crifo1997}. 
    \end{itemize}

    \vspace{0.2cm}
    The spatial density $n_g$ can be correlated after integration to the filling factor parameter of the $Af\rho$ measurement. Considering the \cite{Crifo1997} ejection model, the relation between the local dust sublimation rate and the $Af\rho$ parameter allows us to derive the expression of $K(r_h)$: 
   
   \begin{equation}
    K(r_h)=\frac{J Af\rho(r_h)}{2NQ_{H_2O}(r_h)A(\Phi)}
   \end{equation}
   where $J=\frac{W}{[\alpha A_3(a_1;a_2)+\beta I A_{3.5}(a_1;a_2)]}$, $I=\frac{1}{2}B\left(\frac{3}{4};\frac{1}{2}\right)$ with $B$ the Beta function \citep{Abramowitz1972} and $A_x$ the solution of the integral $\int_{a_1}^{a_2}\frac{da}{a^{u-x+1}}$ \citep{Vaubaillon2005}. The parameter $A(\Phi)$ represents the albedo of the nucleus at the phase angle $\Phi$. \\
   
   The differential local dust sublimation rate is finally 
    \begin{equation}
     Z_g(r_h,\theta,a)=\frac{J Af\rho(r_h)}{2A(\Phi)}\frac{\cos\theta}{\pi R_c^2}\frac{1}{a^u}
    \end{equation} 
  
   \vspace{0.2cm}
   The total number of meteoroids ejected in the size range $[a_1',a_2']$, in a differential region around $r_h$ at the ejection angles $(\theta,\phi)$ during a period $\Delta t$ is 
   
   \begin{equation}\label{eq:int1}
   \begin{aligned}
    N_g=& N_g(r_h,a_1',a_2')\\
    = & R_c^2 \Delta t\int_{\theta-\frac{\Delta \theta}{2}}^{\theta+\frac{\Delta \theta}{2}}  \int_{\phi-\frac{\Delta \phi}{2}}^{\phi+\frac{\Delta \phi}{2}} \int_{r_h-\frac{\Delta r_h}{2}}^{r_h+\frac{\Delta r_h}{2}}
    \\
    &\int_{a_1'}^{a_2'} Z_g(r_h,\theta,a')\sin\theta\mbox{ } da' dr_h d\phi d\theta\\
   \end{aligned}
   \end{equation}
   
   i.e. 
   
   \begin{equation}
   \begin{split}
    N_g&=R_c^2 \Delta t \int_{\theta}\int_{\phi}\int_{r_h}\int_{a'}\frac{J Af\rho(r_h)}{2A(\Phi)}\frac{\cos\theta}{\pi R_c^2} \\
    & \times \frac{1}{a'^u}\sin\theta\mbox{ } da' dr_h d_\phi d\theta\\
    &=\frac{\Delta t J}{2\pi A(\Phi)}\int_{\theta}\sin\theta \cos\theta d\theta \int_{\phi} d\phi \\
    &\int_{r_h} Af\rho(r_h)dr_h \int_{a'} \frac{1}{a'^u}da'
    \end{split}
   \end{equation}
   
   For $\theta \in [0,\frac{\pi}{2}]$ and $\phi \in [0,2\pi]$ we obtain
   
    \begin{equation} \label{eq:2}
    \begin{split}
      N_g=&\frac{J\Delta t}{2 A(\Phi)}A_1(a_1',a_2') \int_{r_h} Af\rho(r_h)dr_h
      \end{split}
    \end{equation}
    
    If we assume that $Af\rho(r_h)=K_1+Af\rho_{max}\cdot10^{-\gamma(r_h-q)}$, with $q$ the perihelion distance and $\gamma=\gamma_1$ pre-perihelion and $\gamma=\gamma_2$ post-perihelion \zcorr{(cf. \ref{sec:comet_photometry}, with \zcorr{$Af\rho_{max}=Af\rho(X_{max})$ and $X_{max}=0$)}}, then 
    
         \begin{equation}
          \begin{split}
          I_{Af\rho}&=\int_{r_h-\frac{\Delta r_h}{2}}^{r_h+\frac{\Delta r_h}{2}} Af\rho(r_h)dr_h\\
          &=K_1\Delta r_h-\frac{\zcorr{Af\rho_{max}}}{\gamma \ln(10)}\left[10^{-\gamma (r_h-q)}\right]_{r_h-\frac{\Delta r_h}{2}}^{r_h+\frac{\Delta r_h}{2}}
          \end{split}
         \end{equation} 
         
    and Equation \ref{eq:2} leads to:
    
          \begin{equation}
          \begin{split}
            N_g&=\frac{J\Delta t}{2 A(\Phi)} A_1(a_1',a_2')\cdot\left(K_1\Delta r_h\color{white}\frac{1}{2}\color{black}\right.\\
            &\left.-\frac{\zcorr{Af\rho_{max}}}{\gamma \ln(10)}\left[10^{-\gamma (r_h-q)}\right]_{r_h-\frac{\Delta r_h}{2}}^{r_h+\frac{\Delta r_h}{2}}\right)
          \end{split}
         \end{equation}

  For a given apparition of the comet with a perihelion distance $q$, the number of particles ejected in all directions in the size bin $[a_1',a_2']$ during $\Delta t$ around $r_h$ is finally: 
    
     \begin{equation} \label{eq:weightsol}
     \left\{
           \begin{aligned}
       N_g=& K_2\frac{J\Delta t}{2 A(\Phi)}A_1(a_1',a_2')\cdot\left(K_1\Delta r_h\color{white}\frac{1}{2}\color{black}\right.\\
       -&\left.\frac{\zcorr{Af\rho_{max}}}{\gamma \ln(10)} \left[10^{-\gamma (r_h+\frac{\Delta r_h}{2}-q)}-10^{-\gamma (r_h-\frac{\Delta r_h}{2}-q)}\right] \right)\\
       \gamma=&\gamma_1 \text{ pre-perihelion and } \gamma=\gamma_2 \text{ post-perihelion}
             \end{aligned}
             \right.
     \end{equation}
     
     The optional $K_2$ proportionality coefficient, constant over all the ejection circumstances, allow to account for systematic biases in the estimate of the intensity of the cometary activity. This parameter, as well as the size distribution index $u$ can be calibrated on previous meteor observations. 
    
\end{document}